\documentclass[twocolumn]{aastex62}
\usepackage{graphicx}
\usepackage[flushleft]{threeparttable}

\shorttitle{ASPECS: molecular line luminosity functions at high redshift}
\shortauthors{Decarli et al.}

\def\Lsun{L$_\odot$}
\def\Msun{M$_\odot$}

\def\Ci{[C\,{\sc i}]}

\def\Cii{[C\,{\sc ii}]}
\def\Ci{[C\,{\sc i}]}

\def\kms{km\,s$^{-1}$}

\def\Kkmspc{K~km\,s$^{-1}$\,pc$^2$}
\def\lsim{\mathrel{\rlap{\lower 3pt \hbox{$\sim$}} \raise 2.0pt \hbox{$<$}}}
\def\gsim{\mathrel{\rlap{\lower 3pt \hbox{$\sim$}} \raise 2.0pt \hbox{$>$}}}

\begin{document}

\title{The ALMA Spectroscopic Survey in the HUDF: Multi-band constraints on line luminosity functions and the cosmic density of molecular gas}

\correspondingauthor{Roberto Decarli}
\email{roberto.decarli@inaf.it}

\author[0000-0002-2662-8803]{Roberto Decarli}
 \affil{INAF --- Osservatorio di Astrofisica e Scienza dello Spazio, via Gobetti 93/3, I-40129, Bologna, Italy}

\author[0000-0002-6290-3198]{Manuel Aravena}
 \affil{N\'{u}cleo de Astronom\'{\i}a, Facultad de Ingenier\'{\i}a y Ciencias, Universidad Diego Portales, Av. Ej\'{e}rcito 441, Santiago, Chile}
\author[0000-0002-3952-8588]{Leindert Boogaard}
 \affil{Leiden Observatory, Leiden University, P.O. Box 9513, NL-2300 RA Leiden, The Netherlands}
\author[0000-0001-6647-3861]{Chris Carilli}
 \affil{National Radio Astronomy Observatory, Pete V. Domenici Array Science Center, P.O. Box O, Socorro, NM 87801, USA}
\author[0000-0003-3926-1411]{Jorge Gonz\'alez-L\'opez}
 \affil{N\'{u}cleo de Astronom\'{\i}a, Facultad de Ingenier\'{\i}a y Ciencias, Universidad Diego Portales, Av. Ej\'{e}rcito 441, Santiago, Chile}
\author[0000-0003-4793-7880]{Fabian Walter}
 \affil{Max Planck Institute for Astronomy, K\"onigstuhl 17, 69117 Heidelberg, Germany}
 \affil{National Radio Astronomy Observatory, Pete V. Domenici Array Science Center, P.O. Box O, Socorro, NM 87801, USA}

\author{Paulo C. Cortes}
 \affil{Joint ALMA Office, Alonso de Cordova 3107, Vitacura, Santiago, Chile}
 \affil{National Radio Astronomy Observatory, Charlottesville, VA 22903, USA}
\author[0000-0003-2027-8221]{Pierre Cox}
 \affil{Institut d'Astrophysique de Paris, Sorbonne Universit\'{e}, CNRS, UMR 7095, 98 bis Blvd. Arago, 75014 Paris, France}
\author{Elisabete da Cunha}
 \affil{The University of Western Australia, ICRAR M468, 35 Stirling Hwy, Crawley WA 6009, Australia}
\author[0000-0002-3331-9590]{Emanuele Daddi}
 \affil{Laboratoire AIM, CEA/DSM-CNRS-Universite Paris Diderot, Irfu/Service d\u2019Astrophysique, CEA Saclay, Orme des Merisiers, F-91191 Gif-sur-Yvette cedex, France}
\author[0000-0003-0699-6083]{Tanio D\'{\i}az-Santos}
 \affil{N\'{u}cleo de Astronom\'{\i}a, Facultad de Ingenier\'{\i}a y Ciencias, Universidad Diego Portales, Av. Ej\'{e}rcito 441, Santiago, Chile}
 \affil{Chinese Academy of Sciences South America Center for Astronomy (CASSACA), National Astronomical Observatories, CAS, Beijing 100101, China}
 \affil{Institute of Astrophysics, Foundation for Research and Technology-Hellas (FORTH), Heraklion, GR-70013, Greece} 
\author{Jacqueline A.~Hodge}
 \affil{Leiden Observatory, Leiden University, P.O. Box 9513, NL-2300 RA Leiden, The Netherlands}
\author{Hanae Inami}
 \affil{Hiroshima Astrophysical Science Center, Hiroshima University, 1-3-1 Kagamiyama, Higashi-Hiroshima, Hiroshima, 739-8526, Japan}
\author[0000-0002-9838-8191]{Marcel Neeleman}
 \affil{Max Planck Institute for Astronomy, K\"onigstuhl 17, 69117 Heidelberg, Germany}
\author{Mladen Novak}
 \affil{Max Planck Institute for Astronomy, K\"onigstuhl 17, 69117 Heidelberg, Germany}
\author[0000-0001-5851-6649]{Pascal Oesch}
 \affil{Department of Astronomy, University of Geneva, Ch. des Maillettes 51, 1290 Versoix, Switzerland}
 \affil{International Associate, Cosmic Dawn Center (DAWN) at the Niels Bohr Institute, University of Copenhagen and DTU-Space, Technical University of Denmark, Copenhagen, Denmark}
\author[0000-0003-1151-4659]{Gerg\"{o} Popping}
 \affil{European Southern Observatory, Karl-Schwarzschild-Strasse 2, 85748, Garching, Germany}
\author[0000-0001-9585-1462]{Dominik Riechers}
 \affil{Cornell University, 220 Space Sciences Building, Ithaca, NY 14853, USA}
\author{Ian Smail}
 \affil{Centre for Extragalactic Astronomy, Department of Physics, Durham University, South Road, Durham, DH1 3LE, UK}
\author{Bade Uzgil}
 \affil{California Institute of Technology, 1200 E. California Boulevard, Pasadena, CA 91125, USA}
\author{Paul van der Werf}
 \affil{Leiden Observatory, Leiden University, P.O. Box 9513, NL-2300 RA Leiden, The Netherlands}
\author{Jeff Wagg}
 \affil{SKA Organization, Lower Withington Macclesfield, Cheshire SK11 9DL, UK}
\author[0000-0003-4678-3939]{Axel Weiss}
 \affil{Max-Planck-Institut f\"ur Radioastronomie, Auf dem H\"ugel 69, 53121 Bonn, Germany}

\begin{abstract}
We present a CO and atomic fine-structure line luminosity function analysis using the ALMA Spectroscopic Survey in the Hubble Ultra Deep Field (ASPECS). ASPECS consists of two spatially--overlapping mosaics that cover the entire ALMA 3\,mm and 1.2\,mm bands. We combine the results of a line candidate search of the 1.2\,mm data cube with those previously obtained from the 3\,mm cube. Our analysis shows that $\sim$80\% of the line flux observed at 3\,mm arises from CO(2-1) or CO(3-2) emitters at $z$=1--3 (`cosmic noon'). At 1.2\,mm, more than half of the line flux arises from intermediate-J CO transitions ($J_{\rm up}$=3--6); $\sim12$\% from neutral Carbon lines; and $< 1$\% from singly-ionized Carbon, \Cii{}. This implies that future \Cii{} intensity mapping surveys in the epoch of reionization will need to account for a highly significant CO foreground. The CO luminosity functions probed at 1.2\,mm show a decrease in the number density at a given line luminosity (in units of $L'$) at increasing $J_{\rm up}$ and redshift. Comparisons between the CO luminosity functions for different CO transitions at a fixed redshift reveal sub-thermal conditions on average in galaxies up to $z\sim 4$. In addition, the comparison of the CO luminosity functions for the same transition at different redshifts reveals that the evolution is not driven by excitation. The cosmic density of molecular gas in galaxies, $\rho_{\rm H2}$, shows a redshift evolution with an increase from high redshift up to $z\sim1.5$ followed by a factor $\sim 6$ drop down to the present day. This is in qualitative agreement with the evolution of the cosmic star--formation rate density, suggesting that the molecular gas depletion time is approximately constant with redshift, after averaging over the star-forming galaxy population.
\end{abstract} 

\section{Introduction} 

Stars form in the dense, molecular phase of the interstellar medium (ISM; see, e.g., reviews in \citealt{kennicutt12}, \citealt{carilli13}, \citealt{dobbs14}, \citealt{combes18}, \citealt{tacconi20}, and \citealt{hodge20}). Molecular gas is thus a key ingredient of galaxy formation, and it plays a critical role in shaping the history of cosmic star formation \citep[e.g.,][]{lilly95,madau96,hopkins06,madau14}. Gauging the amount of molecular gas in galaxies available for star formation, as well as its physical conditions and excitation properties, is thus pivotal in our understanding of the formation and evolution of galaxies. For instance, the cosmic star formation rate density, $\rho_{\rm SFR}$, may result from an evolution of the amount of molecular gas stored in galaxies, averaged over cosmological volume, $\rho_{\rm H2}$, or from an evolution in the efficiency at which molecular gas is converted into stars (as set by the inverse of the depletion time, $t_{\rm dep}$, i.e., the timescale required for the galaxy to exhaust its current gaseous reservoirs, under the assumption that stars keep forming at the current rate), or by a combination of both.

Molecular hydrogen, H$_2$, is a poor radiator \citep[e.g.,][]{omont07}; therefore, observations of the molecular phase of the ISM typically rely on other molecules, in particular the Carbon monoxide, $^{12}$C$^{16}$O (hereafter, CO), which is abundant in the star-forming ISM and efficiently radiates via rotational transitions even at modest excitation energies (corresponding to excitation temperatures of a few 10's K, as observed in the cold, star--forming medium). Low-$J$ CO transitions ($J_{\rm up}\lsim 4$) have rest-frame frequencies, $\nu_0$, of 100--500\,GHz (rest wavelength $\lambda_0$=0.6\,mm--3\,mm), and are often used to gauge the mass in molecular gas, as their luminosity is only modestly dependent on the gas physics (in particular, excitation temperature and density). Intermediate-$J$ CO transitions ($5 \lsim$ $J_{\rm up} \lsim 7$; $\nu_0$=500--900\,GHz, $\lambda_0$=0.3--0.6\,mm) and high-$J$ CO transitions ($J_{\rm up}\gsim 8$, $\nu_0>900$\,GHz), on the other hand, owe their luminosity to the higher excitation, warmer or denser medium -- thus they are better tracers of starbursting activity, nuclear activity, or shocks \citep[see discussions in, e.g.,][]{weiss07,carilli13,daddi15,kamenetzky18,boogaard20}. 

Surveys of molecular gas in high--redshift galaxies are blossoming thanks to the unprecedented observational capabilities offered by the Jansky Very Large Array (VLA), the IRAM NOrthern Expanded Millimeter Array (NOEMA), and the Atacama Large Millimeter Array (ALMA). The number of CO--detected galaxies at $z>0.5$ has increased significantly in the last few years, and now exceeds $250$ \citep[see, e.g., the compilation in][]{tacconi18}. Most of these detections come from targeted investigations, i.e., investigations of the molecular content of known galaxies pre-selected based on their redshift, stellar mass, far--infrared luminosity, star--formation rate (SFR), nuclear activity, apparent luminosity, etc. These studies have been instrumental in effectively establishing empirical relations between gas content and a number of galaxy properties \citep[e.g.,][]{greve05,daddi10a, tacconi10, tacconi13, tacconi18, aravena12, genzel10, genzel11, genzel15, bothwell13, dessaugeszavadsky17}. 

Molecular scans, i.e., interferometric observations of blank fields over a wide frequency range at millimeter wavelengths, represent a powerful complementary approach. By searching for molecular gas emission irrespective of the position and redshift, they effectively result in a line flux--limited survey of a well--defined cosmological volume, and do not depend on any pre--selection. The first molecular scan that reached sufficient depth to secure CO detections in typical galaxies at $z>1$ came from a $>$100-hr long campaign targeting a $\sim 1$\,arcmin$^2$ region of the {\em Hubble} Deep Field North \citep{williams96} in the 3\,mm band using the IRAM/Plateau de Bure Interferometer \citep[PdBI;][]{walter12,walter14,decarli14}. The ALMA Spectroscopic Survey in the {\em Hubble} Ultra Deep Field, ASPECS, built on the success of the PdBI program by performing two frequency scans at 3\,mm and 1.2\,mm. The ASPECS-Pilot program \citep{walter16, aravena16a, aravena16b, decarli16a, decarli16b, bouwens16, carilli16} offered a first glimpse at the molecular gas content in galaxies residing in one of the best--studied regions of the extragalactic sky, the {\it Hubble} Ultra Deep Field \citep{beckwith06}. The ASPECS-Pilot survey was then expanded into an ALMA Large Program (LP) targeting a 4.6\,arcmin$^2$ area, with the same survey strategy \citep{gonzalezlopez19, gonzalezlopez20, decarli19, boogaard19, boogaard20, aravena19, aravena20, popping19, popping20, uzgil19, magnelli20, inami20, walter20}. Among other results, ASPECS provided robust constraints on the low--$J$ CO luminosity functions up to $z\sim 4$, and an estimate of the evolution of the cosmic density of molecular gas in galaxies, $\rho_{\rm H2}$($z$). A follow-up program dubbed VLASPECS used the NSF's Karl G. Jansky Very Large Array, VLA, to secure 30.6--38.7 GHz coverage over part of the ASPECS footprint, thus providing a low--$J$ anchor to CO excitation models for galaxies by directly measuring CO(1-0) luminosities in the redshift range $z$=2.0--2.7 \citep{riechers20}.

Other molecular scan efforts appeared in the literature in the last couple of years: The COLDz survey used $>$320\,hr of the VLA time to sample CO(1-0) emission at $z\approx2-3$ (`cosmic noon') as well as CO(2-1) at $z\approx5-7$ over $\sim 60$\,arcmin$^2$ in parts of the COSMOS \citep{scoville07} and GOODS-North \citep{giavalisco04} fields \citep{pavesi18,riechers19,riechers20b}. \citet{lenkic20} used the Plateau de Bure High-z Blue-Sequence Survey 2 (PHIBSS2) data \citep{tacconi18} to search for serendipitous emission in the cubes, besides the central targets. 
These studies place first direct constraints on the CO luminosity function in galaxies at $z\sim 2$, and revealed a higher molecular content in galaxies at these redshifts compared to the local universe: $\rho_{\rm H2}(z=2-3)\approx (1-20)\times 10^7$\,\Msun{}\,Mpc$^{-3}$.
A few serendipitous molecular line detections have been reported in the the fields of sub-millimeter galaxies \citep{wardlow18,cooke18}, in an ALMA deep field around SSA22 \citep{hayatsu17} and around graviational lensing clusters \citep{yamaguchi17,gonzalezlopez17}.
Finally, \citet{klitsch19} used the high signal--to--noise ratio (S/N) of mm-bright calibrators in the ALMA archive to search for CO absorption features. They do not detect any extragalactic source, which sets constraints on both the CO luminosity functions and $\rho_{\rm H2}$($z$) up to $z\sim 1.7$. In addition to CO--based estimates, various studies have inferred molecular gas mass functions and $\rho_{\rm H2}$($z$) via estimates based on the dust continuum, but this relies on an empirically--calibrated gas-to-dust conversion \citep[e.g.,][]{scoville17, liu19, magnelli20}.

In this paper, we capitalize on the completed ASPECS dataset in order to constrain the luminosity functions and average cosmic content of molecular gas in galaxies throughout cosmic time. First, we present the new 1.2\,mm dataset (\S~\ref{sec_obs}), the ancillary data (\S~\ref{sec_ancillary}), and the approach adopted in the analysis (\S~\ref{sec_results}). Then, we complement the 1.2\,mm dataset with the information from the 3\,mm part of ASPECS in a homogeneous analysis of molecular and atomic line emission from the cold ISM in high-redshift galaxies (\S~\ref{sec_discussion}). We present our conclusions in \S~\ref{sec_conclusions}. Throughout this paper we adopt a $\Lambda$CDM cosmological model with $H_0=70$ km\,s$^{-1}$\,Mpc$^{-1}$, $\Omega_{\rm m}=0.3$ and $\Omega_{\Lambda}=0.7$ \citep[consistent with the measurements by the][]{planck15}. 

\begin{figure*}
\begin{center}
\includegraphics[width=0.99\columnwidth]{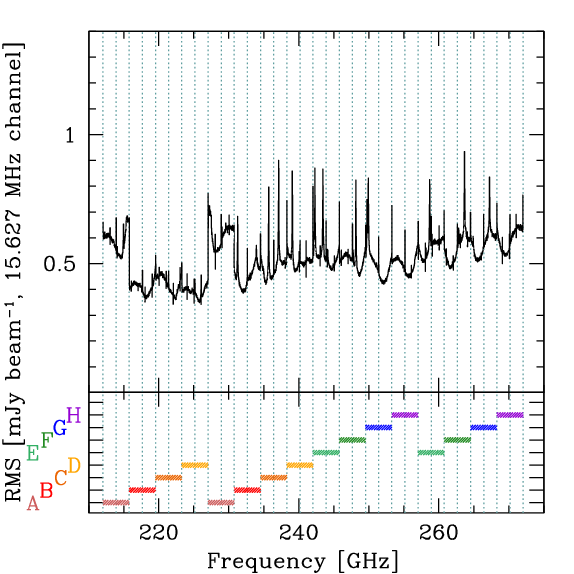}
\includegraphics[width=0.99\columnwidth]{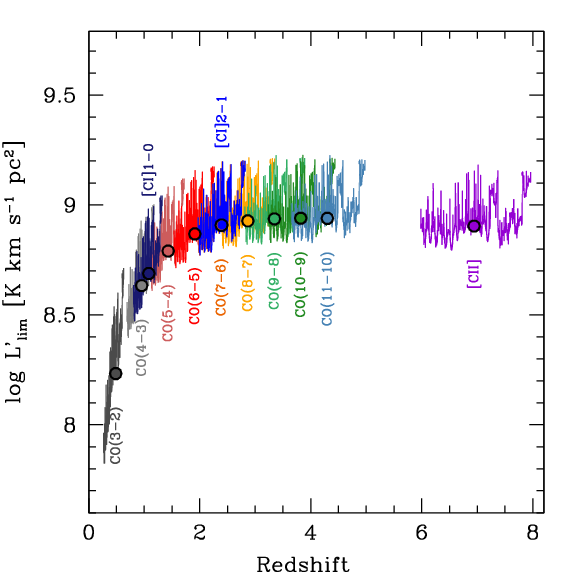}\\
\end{center}
\caption{Sensitivity limits of the ASPECS 1.2\,mm cube. {\em Left:} Channel rms as a function of frequency. For a 15.6 MHz channel, the typical rms is $\sim$0.5\,mJy\,beam$^{-1}$ throughout the entire band. The frequency settings used in the observations (labeled A--H) and the edges of each spectral window are also marked. {\em Right:} Line luminosity limits (in units of \Kkmspc{}) as a function of redshift. Here we assume a 5-$\sigma$ limit for a line width of 200\,\kms{}. The dots highlight the fiducial limit, obtained as the median sensitivity throughout the band.}
\label{fig_limits}
\end{figure*}

\begin{table*}
\caption{\rm Emission lines, corresponding redshift bins, volume--weighted average redshift, cosmic volume (in comoving units, within the area of $>50$\% sensitivity), and typical 5-$\sigma$ line luminosity limit at $\langle z \rangle$, assuming a line width of 200\,\kms{} in ASPECS LP 1.2\,mm (observed range: 212--272\,GHz).} \label{tab_zbin}
\begin{center}
\begin{tabular}{cccccc}
\hline
Line    & Redshift      & $\langle z \rangle$ & Volume     & limit $L$        & limit $L'$       \\
        &               &            &[Mpc$^3$] & [$10^7$\,\Lsun]  & [$10^8$\,\Kkmspc] \\
 (1)    &     (2)       & (3)        & (4)              &  (5)      & (6)       \\
\hline
   CO(3--2)    & $0.2711-0.6306$ &0.49&	 921.3 &   0.023  & 1.710  \\
   CO(4--3)    & $0.6947-1.1740$ &0.96&	2960.9 &   0.135  & 4.299  \\
   CO(5--4)    & $1.1183-1.7173$ &1.43&	5106.3 &   0.378  & 6.174  \\
   CO(6--5)    & $1.5418-2.2606$ &1.91&	6923.8 &   0.781  & 7.384  \\
   CO(7--6)    & $1.9651-2.8037$ &2.39&	8470.4 &   1.358  & 8.088  \\
   CO(8--7)    & $2.3884-3.3467$ &2.87&	9597.2 &   2.121  & 8.464  \\
   CO(9--8)    & $2.8115-3.8895$ &3.35& 10478.0 &   3.085  & 8.647  \\
  CO(10--9)    & $3.2345-4.4321$ &3.82& 11012.3 &   4.262  & 8.712  \\
 CO(11--10)    & $3.6574-4.9745$ &4.30& 11371.6 &   5.660  & 8.696  \\
\hline				 	    		   
~[CI]$_{1-0}$ & $0.8091-1.3207$ &1.08&	3540.8 &   0.186  & 4.878  \\
~[CI]$_{2-1}$ & $1.9750-2.8164$ &2.40&	8509.3 &   1.374  & 8.102  \\
\hline				 	    		   
    [CII]     & $5.9861-7.9619$ &6.94& 12621.6 &  17.61   & 8.018  \\
\hline
\end{tabular}
\end{center}
\end{table*}

\section{Observations}

\subsection{ALMA data}\label{sec_obs}

The ASPECS LP survey is an ALMA Cycle 4 Large Program comprising two bands, at 3\,mm and 1.2\,mm. The former is presented and discussed elsewhere \citep{gonzalezlopez19, decarli19, boogaard19, aravena19, popping19, uzgil19, inami20}. The latter consists of a mosaic of 85 pointings in the eXtremely Deep Field (XDF, \citealt{illingworth13}; also dubbed {\it Hubble} Deep Field 2012 or HUDF12, \citealt{koekemoer13}) for a total area of 4.2 arcmin$^2$ down to 10\% sensitivity, or 2.9 arcmin$^2$ within the 50\% primary beam response. The observing strategy involves covering the full mosaic area at each telescope visit. The pointings were arranged in classical hexagonal patterns at $11''$ separation, which ensures Nyquist sampling throughout the entire frequency range of the observations and results in a spatially--uniform sensitivity throughout the majority of the footprint.

Observations were carried out in two parts, a first pass in 2017, March -- April (roughly 20\% of the total data volume spread among all of the requested frequency settings) and the remainder in 2018, May -- July. The 2017 observations were collected with average weather conditions, with precipitable water vapour $2.5$--$3.0$\,mm; on the other hand, the 2018 observations were gathered in excellent weather conditions, with precipitable water vapour $\sim 0.6$\,mm in most of the executions. The array was in compact, C40-1 or C40-2 configurations, with baselines in the range 15\,m--320\,m.

The observations sampled eight different frequency tunings, continuously encompassing the entire 212--272\,GHz window (see Fig.~\ref{fig_limits}). Quasars J0329--2357, J0334--4008, J0348--2749, and J0522--3627 were employed as pointing, phase, amplitude, and bandpass calibrators. 

We processed the raw data using the {\sc casa} calibration pipeline for ALMA \citep[v.5.1.1; see][]{mcmullin07}. No additional flagging was applied. We inverted the visibilities using the task \textsf{tclean}, and adopting natural weighting. The resulting beam is $\sim1.5''\times1.1''$. Along the spectral axis, the cube was resampled using 15.627\,MHz wide channels ($\approx 19$\,\kms{} at 242 GHz). Cleaning was performed down to 2-$\sigma$ per channel after putting cleaning boxes on all the sources with S/N$>$5 in their continuum emission. We reach a sensitivity of $\sim 0.5$\,mJy\,beam$^{-1}$ per 15.627\,MHz channel roughly constant throughout the 1.2\,mm band (see Fig.~\ref{fig_limits}). We also created a continuum--subtracted version of the cube, after identifying and excluding the channels with the brightest emission lines (see \citealt{gonzalezlopez20} for details). Finally, we created a tapered version of the cube, where we degrade the angular resolution by setting the \textsf{restoringbeam}=$2''$ in the task \textsf{tclean}. We use this tapered cube to extract 1D spectra of the detected galaxies, following \citet{gonzalezlopez19}.

\subsection{Ancillary data}\label{sec_ancillary}

The targeted field lies in the {\em Hubble} Ultra Deep Field (HUDF), arguably the best studied extragalactic field in the sky. We employ the 3D-{\em HST} photometric catalog by \citet{skelton14}, which relies on optical {\em Hubble}/Advanced Camera for Surveys data \citep{beckwith06}, deep near-infrared {\em Hubble}/Wide Field Camera 3 observations from the Cosmic Assembly Near-infrared Deep Extragalactic Legacy Survey (CANDELS; \citealt{grogin11}; \citealt{koekemoer11}), enriched with multi-wavelength photometry and spectroscopy from various surveys \citep[see][and references therein]{boogaard19}. In particular, the MUSE {\em Hubble} Ultra Deep Survey \citep{bacon17} provides integral field spectroscopy of a $3'\times 3'$ field (encompassing the whole HUDF) over the wavelength range 4750--9300 \AA{}. More than $1500$ galaxies have secured redshifts from MUSE \citep{inami17}, $\sim 700$ of which are within the area of the ASPECS LP 1.2\,mm mosaic with $>$50\% primary beam response.

When comparing ALMA observations to other catalogs, we account for a known systematic astrometry offset ($\Delta$RA=$+0.076''$, $\Delta$Dec=$-0.279''$) between optical and mm/radio data \citep{rujopakarn16,dunlop17}.

\section{Analysis and Results}\label{sec_results}

\subsection{Line search at 1.2\,mm}\label{sec_search}

We search for emission lines in the original and the continuum-subtracted ASPECS LP 1.2\,mm cubes using \textsc{findclumps} \citep{walter16,decarli19,gonzalezlopez19}. The code performs a floating average of channels over various kernel widths (with one channel corresponding to $\approx 19$\,\kms{} at the center of the bandwidth). Each averaged channel is searched for both positive and negative peaks. The S/N of a line candidate is computed as the ratio between the flux density measured at the centroid of the line candidate and the rms of the map used in the line identification. We refer to the line search results from the continuum--subtracted cube for line candidates that lie within $2''$ from a bright continuum source from the compilation in \citet{gonzalezlopez20}, and to the results from the original cube for anywhere else in the mosaic. 

\begin{figure}
\begin{center}
\includegraphics[width=0.99\columnwidth]{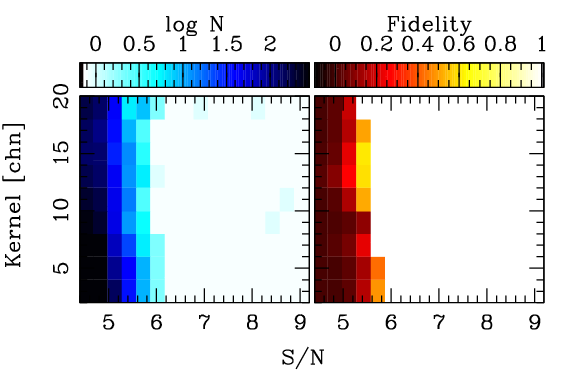}\\
\includegraphics[width=0.99\columnwidth]{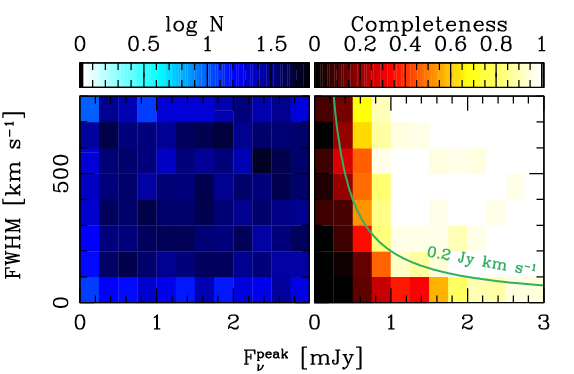}\\
\end{center}
\caption{{\em Top-left:} Number of observed (positive) line candidates from the line search as a function of S/N and kernel width that maximized the S/N of the line candidate in the line search. {\em Top-right:} Best-fits of the fidelity dependence on S/N and kernel width. The fidelity of line candidate is close to unity at S/N$>$5.8, and drops rapidly to zero at S/N$<$5. The fidelity at a given S/N increases with increasing line widths, as expected because of the fewer independent noise realizations in the cubes. {\em Bottom-left:} Number of simulated lines injected in the cube for completeness assessments, as a function of line peak flux, $F_{\nu}^{\rm peak}$, and width (parametrized as FWHM). Only lines located within the footprint at $>$50\% response in the cube. {\em Bottom-right:} Completeness of the line search. The completeness is $\gsim$90\% for virtually any line with integrated flux larger than 0.2\,Jy\,\kms{} (indicated by a green solid line) and peak fluxes of $>$1\,mJy.}
\label{fig_compl_fidelity}
\end{figure}

Positive peaks are a combination of signal from astrophysical sources and noise, while negative peaks are only due to noise. 
The latter are thus used to statistically infer the reliability or `fidelity' of a line candidate, given its width ($\sigma_{\rm line}$) and signal to noise (S/N):
\begin{equation}\label{eq_fidelity}
{\rm Fidelity(S/N,\sigma_{\rm line})}=1 - \frac{N_{\rm neg}({\rm S/N,\sigma_{\rm line}})}{N_{\rm pos}({\rm S/N,\sigma_{\rm line}})}
\end{equation}
where $N_{\rm pos,neg}$ is the number of line candidates in a given S/N and $\sigma_{\rm line}$ bin. Only S/N$>$4 line candidates are considered in this analysis. For each line width bin, we fit the observed distribution of the noise peaks with the tails of a Gaussian function centered at zero, and the additional signal due to real sources as a power law. The fit is performed in two steps, first by modeling the negative distributions in $\sigma_{\rm line}$ bins, then by fitting the positive distributions capitalizing on the posterior parameters of the negative fits for the noise component of the observed distributions. This allows us to mitigate limitations due to the low number of entries in some bins, while properly accounting for their statistical relevance. Following \citet{pavesi18}, \citet{gonzalezlopez19}, and \citet{decarli19}, we conservatively treat these estimates of the fidelity as upper limits; e.g., in each realization of the luminosity functions, a line with a fidelity of 40\% has up to 40\% chance to be used in the analysis. The upper--right panel of Fig.~\ref{fig_compl_fidelity} show the behaviour of the fidelity as a function of the adopted kernel width (i.e., the number of channels that maximizes the S/N of a line candidates -- this is a proxy of the line width) as well as of the integrated S/N of the line candidate. The fidelity is close to 100\% for any line at S/N$>$6, and drops rapidly to zero between S/N=5--6, with narrower lines being typically less reliable than broader lines with the similar total S/N. We refer the reader to \citet{decarli19} and \citet{gonzalezlopez19} for detailed discussions on the assessment of the line reliability. Finally, we adopt a fidelity of unity (not treated as an upper limit) for the high--significance line candidates associated with known sources for which we have clear 1.2\,mm continuum counterparts, as well as a spectroscopic redshift from MUSE or from our 3\,mm line search. These sources are studied in detail in \citet{boogaard20} and \citet{aravena20}. The final catalog from the line search consists of 234 line candidates with fidelity $>0.2$, 75 with fidelity $>0.5$, and 35 with fidelity $>0.8$.

We estimate the completeness by injecting simulated emission lines with a range of input parameters into the observed data cube. We adopt a 3D Gaussian profile for mock lines. In the spatial dimension, we assume the position angle and width of the major and minor axes of the synthesized beam (i.e., sources are spatially unresolved). We run the line search on the cube, and then define the completeness as a function of the input line parameters as the ratio between the number of retrieved versus injected sources. As input parameters, we consider the Right ascension, $\alpha$; the declination, $\delta$; the observed frequency, $\nu_{\rm obs}$; the line width along the spectral axis, FWHM=$2\sqrt{2\,\ln 2} \,\sigma_{\rm line}$; the line peak intensity, $F_\nu^{\rm peak}$. Sources are distributed uniformly in the sampled parameter space (corresponding to the actual 3D coverage of ASPECS LP 1.2\,mm mosaic in terms of $\alpha$, $\delta$, and $\nu_{\rm obs}$; and ranging between 0--800\,\kms{} and 0--3\,mJy in terms of FWHM and $F_\nu^{\rm peak}$). 
A total of 8000 mock lines were injected, $>3000$ of which reside within the area with $>50$\% primary beam response. The bottom panels of Fig.~\ref{fig_compl_fidelity} show the number of injected lines as a function of FWHM and $F_\nu^{\rm peak}$, and the associated completeness in bins of 100\,\kms{} and 0.25\,mJy in line width and peak flux. The other free parameters in our simulation do not appear to significantly affect the completeness of the line search (after accounting for the primary beam response). We drop all line candidates with a completeness of $<0.2$ from our analysis. The median correction due to completeness is $<30$\%.

\subsection{Line fluxes}

For each line candidate, we extract a 1-D spectrum from the pixel where the line spatial centroid is found. We then fit the extracted spectrum with a continuum and a Gaussian profile, using our custom Bayesian Monte Carlo Markov Chain procedure, using \textsc{findclumps} results as priors \citep[see][]{decarli19}. 

\begin{figure}
\begin{center}
\includegraphics[width=0.99\columnwidth]{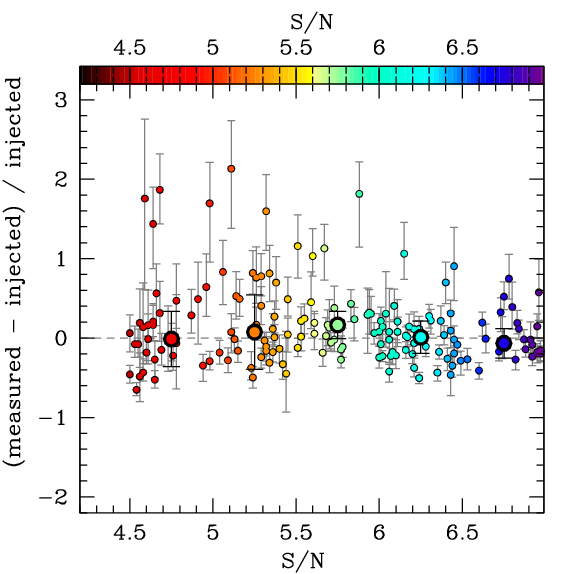}\\
\end{center}
\caption{The impact of flux boosting on our analysis, estimated by comparing the injected and measured fluxes of mock lines. Small symbols show a random subset of individual mock lines, larger symbols are median values in bins of $\Delta$S/N=0.5. Flux boosting affects the flux measurement of $\approx 10$\% of lines at S/N$<$5, and is completely negligible at S/N$>$6.}
\label{fig_flux_boost}
\end{figure}

As we push our search towards the detection limit of our survey, we might tend to preferentially pick sources that appear brighter than they are due to noise fluctuations. We investigate the impact of flux boosting by comparing the injected and recovered fluxes of mock lines (see Sec.~\ref{sec_search} for details on the line simulations). Fig.~\ref{fig_flux_boost} compares the measured versus injected fluxes as a function of the detection S/N. The measured flux is typically within 30\% of the input flux (at 1-$\sigma$) in the $4.5<$ S/N $<7$ regime. Flux boosting appears to be significant (i.e., the recovered flux exceeds 3-$\sigma$ of the distribution width)\footnote{The impact of flux boosting is likely larger for spatially--extended sources \citep[see, e.g.][]{pavesi18}. However, our analysis assumes unresolved emission in the tapered cube for all of the sources.} in $\approx 10$\% of sources with S/N$<$5, and $\sim 1$\% of the sources at S/N$>$6. Because of the modest fidelity of sources with S/N$<$5.8, we consider flux boosting negligible for the purpose of our analysis.

\subsection{Line identification and redshifts}

\subsubsection{Sources with a near-infrared counterpart}

\begin{figure}
\begin{center}
\includegraphics[width=0.99\columnwidth]{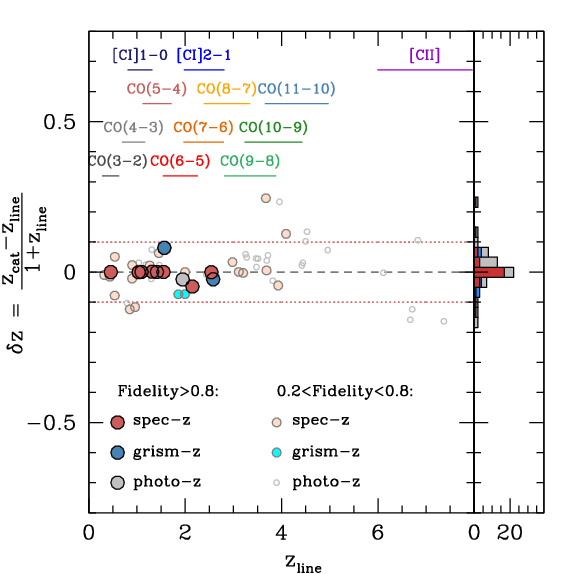}\\
\end{center}
\caption{Redshift match from the line search in the ASPECS LP 1.2\,mm mosaic, $z_{\rm line}$, and the ancillary catalog values, $z_{\rm cat}$, based on the 3D-HST catalog \citep{skelton14}, augmented with the most up to date spectroscopic information (see Sec.~\ref{sec_ancillary} for details). We consider good matches cases where $|\delta z|<0.1$ ($<0.01$ for sources with spectroscopic redshifts). All of the high fidelity lines have a matching redshift in the catalog. The redshift ranges mapped by the various transitions considered in this work are marked as horizontal bars.}
\label{fig_z_match}
\end{figure}

Table~\ref{tab_zbin} lists the transitions we are sensitive to, in various redshift bins\footnote{The ASPECS LP 1.2\,mm coverage formally includes also the CO(2-1) transition at $z<0.0874$. However, the sampled volume within this redshift range is $\sim 3.9\times10^{-4}$\,Mpc$^3$, insufficient for this analysis.}. In order to identify the rest-frame transition associated with a given line candidate, we first cross--match our line candidate compilation with catalogs from ancillary data (see Sec.~\ref{sec_ancillary}). All the entries in our galaxy catalog have a redshift estimate (with a wide range of accuracy, from very high for MUSE--identified sources with several bright emission lines to very poor for faint, photometric dropouts detected only in a handful of broad band filters). For each line candidate, we consider as potential counterpart sources within $1''$ from the line spatial centroid. We identify the transition as the one that would yield the closest line redshift, $z_{\rm line}$, to the one reported in the ancillary catalog, $z_{\rm cat}$. We consider good matches line candidates that are found within $1''$ from a known optical/near-infrared counterpart, and with a redshift separation of $|\delta z|=|z_{\rm cat}-z_{\rm line}|/(1+z_{\rm line}) < 0.1$ ($0.01$ for sources with a spectroscopic redshift). All of the fidelity$>$0.8 lines in the search have a clear counterpart (see Fig.~\ref{fig_z_match}). 

Ignoring the effects of gravitational lensing, we can estimate the impact of chance associations (i.e., the probability of intersecting a galaxy at a random point in our datacube) as:
\begin{equation}\label{eq_probchance}
P({\rm chance})=\sum_i \,\frac{A_{\rm beam}}{A_{\rm footprint}}\,\frac{2 \sigma_z}{(1+z) \, \Delta z_i}
\end{equation}
where $A_{\rm beam}$ and $A_{\rm footprint}$ are the areas of the synthesized beam and of the ASPECS LP 1.2\,mm footprint, respectively; $\sigma_z$ is the uncertainty in the redshift, which we assume to be 0.1; $\Delta z_i$ is the redshift coverage of ASPECS LP 1.2\,mm in transition $i$; and the index $i$ runs through the various transitions considered in our analysis. After summing over all of the transitions, we find that the probability of chance association is $\sim 4.3$\%, i.e., from all the line candidates with a counterpart entering our analysis, only a handful of chance associations are expected (and virtually zero if one considers spectroscopic redshift uncertainties instead).

\begin{figure*}
\begin{center}
\includegraphics[width=0.99\columnwidth]{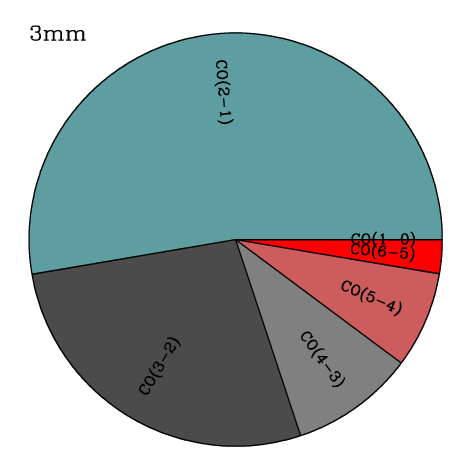}
\includegraphics[width=0.99\columnwidth]{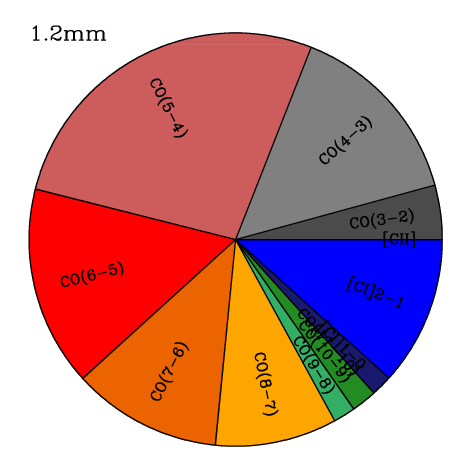}\\
\end{center}
\caption{Pie charts of the fidelity--corrected flux distribution of the lines detected in the ASPECS LP 3\,mm ({\em left}) and 1.2\,mm ({\em right}) cubes. The 3\,mm cube is dominated by low--$J$ CO transitions observed at $z$=1--3. At 1.2\,mm, about 62\% of the total line emission arises from CO transitions with intermediate $J$=3--6. Higher-$J$ transitions account for 25\% of the total line flux in the 1.2\,mm band. The two \Ci{} lines account for $\sim 12$\% of the total line emission, while \Cii{} contributes $<$1\%.}
\label{fig_flux_distr}
\end{figure*}

Fig.~\ref{fig_flux_distr} shows a pie chart of the fidelity--corrected total flux of all the line candidates with an optical/near-infrared counterpart and with fidelity $>$0.5. 
The 3\,mm flux distribution is dominated by CO(2-1) (53\%) and CO(3-2) (27\%), observed at $z$=1--3, while higher-$J$ lines contribute progressively less [CO(4-3): 10\%; CO(5-4): 7\%; CO(6-5): 3\%]. On the other hand, more than half of the total flux measured in lines (62\%) in the ASPECS LP 1.2\,mm mosaic comes from intermediate--J CO transitions (3$\leq$$J_{\rm up}$$\leq$6); 25\% arises from higher-$J$ CO transitions; 12\% from \Ci{}; and less than 1\% from \Cii.  The uncertainties on these fractions are of $\sim 25$\% for the CO lines, and $\sim 50$\% for the Carbon lines, as estimated from the Poissonian uncertainties. The fact that the contribution of \Cii{} flux to the total line flux is $<$1\% in band 6 implies significant challenges for intensity mapping experiments of \Cii{} emission in the epoch of reionization  \citep[e.g.,][]{crites14,lagache18,sun18,yue15,yue19,chung20} as the signal will be dominated by CO foreground emission.

\subsubsection{Sources without a near-infrared counterpart}

The identification of lines without an optical / near-infrared counterpart (roughly 1/3 of the line candidates with fidelity $>$0.8) is done via a bootstrap approach, following, e.g., \citet{decarli19}. 
Here we assume that the probability distribution of a line identification is proportional to the volume sampled in each transition, scaled by a weight set to be equal to $r_{J1}=(0.46, 0.25, 0.12, 0.04)$ for $J_{\rm up}=(3, 4, 5, 6)$, and to $0.01$, 0.05 and 0.003 for $J_{\rm up}>6$, \Ci{}$_{1-0}$ and \Ci{}$_{2-1}$, respectively. These weights have been derived from the average CO spectral energy distribution derived for the ASPECS sources by \citet{boogaard20}. The weights for \Ci{} lines are defined based on a fiducial flux ratio between \Ci{} lines and neighboring CO transitions \citep[see, e.g.,][]{walter11,boogaard20}. Finally, we do not include \Cii{}, based on the flux distribution shown in Fig.~\ref{fig_flux_distr} and the analyses presented in \citet{uzgil20} and \citet{loiacono20}. We discuss different choices of assigning CO transtions (i.e., redshifts) to sources with no near-infrared counterpart in Appendix \ref{sec_probs}.

\subsection{Line luminosities and molecular gas masses}\label{sec_measures}

\begin{figure}
\begin{center}
\includegraphics[width=0.99\columnwidth]{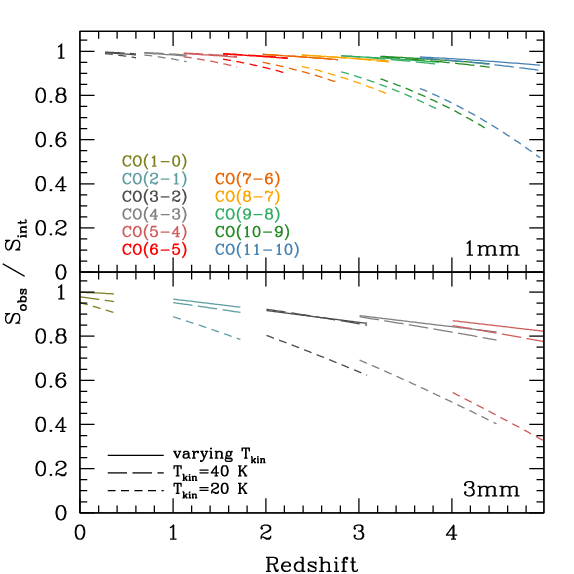}\\
\end{center}
\caption{The effect of the CMB on the observed line fluxes. The correction is computed under the assumption of local thermal equilibrium, for different values of the gas kinetic temperature, $T_{\rm kin}$=20\,K, 40\,K, and for a redshift-dependent description as in \citet{magnelli14}. The correction is always $\lsim 20$\% up to $z\sim 3$ for any $T_{\rm kin}>20$\,K, and $<10$\% up to $z\sim 5$ for any $T_{\rm kin}>40$\,K.}
\label{fig_cmb}
\end{figure}

Line fluxes are transformed into luminosities following, e.g., \citet{carilli13}:
\begin{equation}\label{eq_linelum}
\frac{L'}{\rm K\,km\,s^{-1}\,pc^2}=\frac{3.257\times10^7}{1+z} \, \frac{F_{\rm line}}{\rm Jy\,km\,s^{-1}} \, \left(\frac{\nu_{\rm 0}}{\rm GHz}\right)^{-2} \left(\frac{D_{\rm L}}{\rm Mpc}\right)^2
\end{equation}
where $F_{\rm line}$ is the integrated line flux, $\nu_0$ is the rest-frame frequency of the line, and $D_{\rm L}$ is the luminosity distance. We also compute line luminosities in solar units as:
\begin{equation}\label{eq_linelum2}
\frac{L}{\rm L_\odot}=\frac{1.04\times10^3}{1+z} \, \frac{F_{\rm line}}{\rm Jy\,km\,s^{-1}} \, \frac{\nu_{\rm 0}}{\rm GHz} \left(\frac{D_{\rm L}}{\rm Mpc}\right)^2.
\end{equation}
As our observations probe the rest-frame far-infrared wavelengths of high--redshift galaxies, the Cosmic Microwave Background (CMB) might have an impact on the observed line fluxes. It provides an extra contribution to excitation temperature of the lines, but it also represents a background against which sources are observed. We follow the formalism presented in \citet{dacunha13} to compute the correction between the observed versus intrinsic line fluxes. The correction depends on the intrinsic excitation temperature in the gas. Here we assume local thermal equilibrium ($T_{\rm kin}=T_{\rm exc}$). Fig.~\ref{fig_cmb} shows the correction terms for two fixed temperature values $T_{\rm kin}=20$, 40\,K, and for a redshift-dependent $T_{\rm kin}$ following the dust temperature evolution presented in \citet{magnelli14}. We find that the correction is always $<$20\% for any temperature of interest $T_{\rm kin}>20$\,K, up to $z\sim 3$, and $<$10\% for any $T_{\rm kin}>40$\,K, for all the 1.2\,mm CO lines. In Fig.~\ref{fig_cmb} we also show that the correction would be larger for lines observed at 3\,mm, but still $<$20\% at any $z\lsim 4$ for $T_{\rm kin}$=40\,K. Because the exact correction depends on the (unknown) excitation temperature of the gas in our sources and on the (unverified) validity of the local thermal equilibrium, and given how small the corrections are, we opt not to apply any CMB--related correction in the remainder of our analysis.

The lower--$J$ CO transitions are converted into CO(1-0) luminosities by adopting the CO[$J$-($J$-1)]--to--CO(1-0) luminosity ratios, $r_{J1}$, from the analysis of the CO excitation in CO--detected galaxies in ASPECS LP by \citet{boogaard20}: $L'$ [CO(1-0)] = $L' / r_{J1}$, with $r_{J1}=\{0.75\pm0.11$, $0.46\pm0.07$, $0.31\pm0.07\}$, for $J_{\rm up}$=$\{$2, 3, 4$\}$. We also correct the results from ASPECS LP 3\,mm \citep{decarli19} accordingly for galaxies at $z<2$. At higher redshifts, we adopt $r_{J1}=\{0.80\pm0.14$, $0.61\pm0.13\}$, for $J_{\rm up}$=$\{$3, 4$\}$. As discussed in \citet{boogaard20}, the redshift dependence reflects the higher IR luminosity and IR surface brightness in the higher-redshift ASPECS LP sample \citep[see also][]{aravena20}. We are consistent within uncertainties with the measurements of individual sources. As in \citet{decarli19}, we include bootstrapped realizations of the uncertainties on $r_{J1}$ in the conversion. 

Finally, the CO(1-0) luminosities are converted into corresponding H$_2$ mass: $M_{\rm H2} = \alpha_{\rm CO} \, L'_{\rm CO(1-0)}$ \citep[see][for a review]{bolatto13}. The bulk of the flux emission in our observations arises from typical galaxies with close--to--solar metallicity \citep{boogaard19,aravena19,aravena20}, for which a Galactic conversion factor should apply. Following the literature consensus, we adopt $\alpha_{\rm CO}$=3.6 \Msun{}\,(\Kkmspc)$^{-1}$ \citep[e.g.,][]{daddi10a}. All the results based on $\alpha_{\rm CO}$ would scale linearly if a different (but constant) value is adopted. 

Atomic Carbon transitions can also be used to infer constraints on the gas mass \citep[see, e.g.,][]{weiss05, walter11, alaghbandzadeh13, bothwell17, popping17, valentino18}. In the assumption of optically--thin line emission, the luminosity of the two \Ci{} transitions is related to the mass in neutral Carbon as follows:
\begin{equation}\label{eq_Mci10}
M_{\rm CI}/{\rm M_\odot}=5.706 \times 10^{-4}\,\frac{Q_{\rm ex}}{3} \, e^{23.6/T_{\rm ex}}
\, L'_{\rm [CI]1-0}
\end{equation}
\begin{equation}\label{eq_Mci21}
M_{\rm CI}/{\rm M_\odot}=5.273 \times 10^{-3} \, \frac{Q_{\rm ex}}{5} \, e^{62.5/T_{\rm ex}} \, L'_{\rm [CI]2-1}
\end{equation}
where $Q_{\rm ex}$=$1+3\,e^{-23.6/T_{\rm ex}}+5\,e^{-62.5/T_{\rm ex}}$ is the partition function, $T_{\rm ex}$ is the excitation temperature in K, and line luminosities are quoted in units of \Kkmspc{}. The mass estimates in equations \ref{eq_Mci10} and \ref{eq_Mci21} can be related to the molecular gas mass, under the assumption that all of the Carbon is in neutral form. Assuming an abundance ratio X[C{\sc i}]/X[H$_2$]=$1.9\times10^{-5}$ (\citealt{boogaard20}, consistent with the $10^{-4.8\pm0.2}$ value reported by \citealt{valentino18}), we obtain $M_{\rm H2}=M_{\rm CI}/ (6\, {\rm X[CI]/X[H_2]})$, where the factor of six accounts for the mass ratio between molecular Hydrogen and the Carbon atom. In our analysis, we assume $T_{\rm ex}=29\pm6$\,K \citep{walter11}. 

The \Ci{} transitions have a number of advantages as molecular gas masses. In particular, $M_{\rm CI}$ in Eq.~\ref{eq_Mci10} is nearly linear with $L'_{\rm [CI]1-0}$ for $T_{\rm ex}\gsim 15$\,K (a realistic scenario at high redshift), and optical depth is virtually never an issue once averaged over galactic scales. In principle, the mass estimates inferred via Eq.~\ref{eq_Mci10} and \ref{eq_Mci21} are lower limits on $M_{\rm H2}$, because of the assumption that all of the Carbon is in neutral form; however, the same assumption is usually at the root of the abundance estimates, i.e., the uncertainty cancels out. An additional caveat to consider is that, because \Ci{} is mostly optically thin, these \Ci{}--based mass estimates are more sensitive to assumptions on Carbon abundance and to the fraction of \Ci{} emitted from the neutral versus molecular medium than CO--based estimates.

\begin{figure*}
\begin{center}
\includegraphics[width=0.24\textwidth]{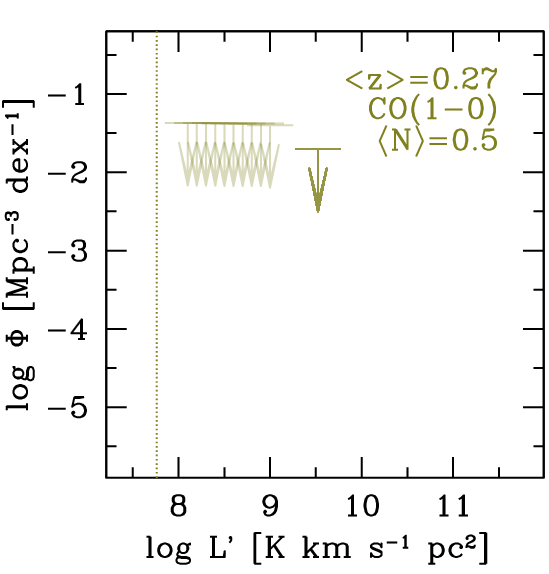}
\includegraphics[width=0.24\textwidth]{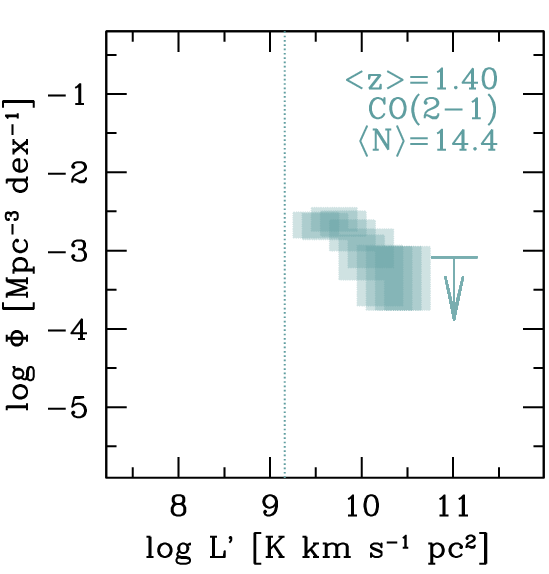}
\includegraphics[width=0.24\textwidth]{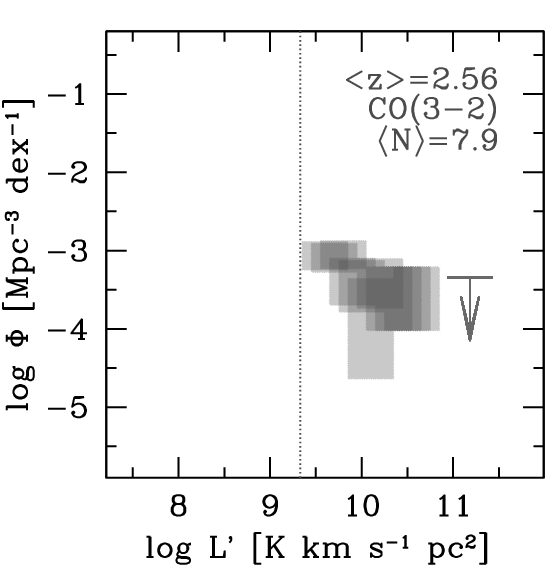}
\includegraphics[width=0.24\textwidth]{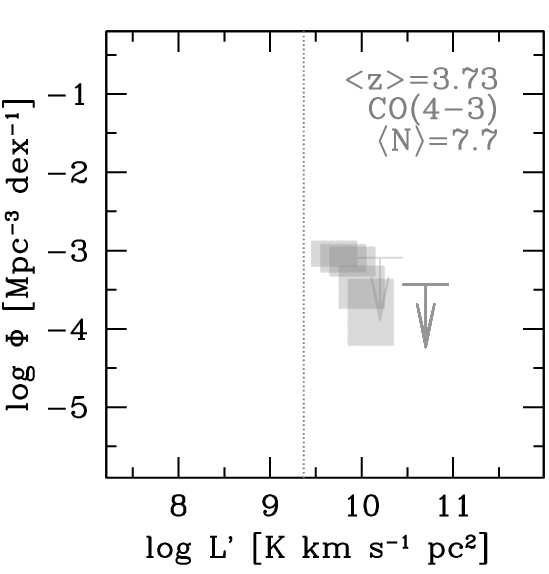}\\
\includegraphics[width=0.24\textwidth]{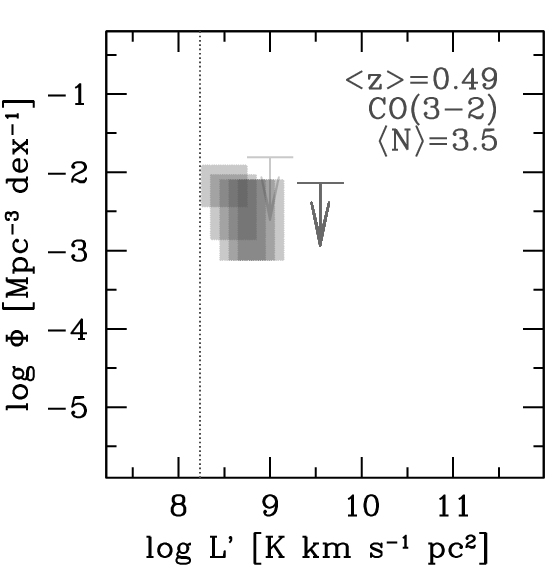}
\includegraphics[width=0.24\textwidth]{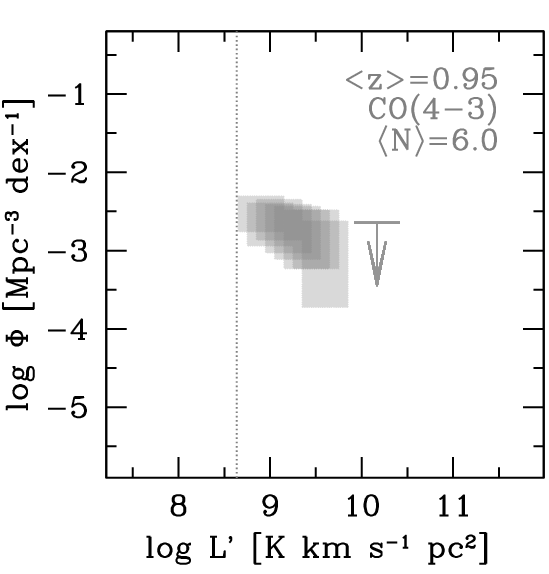}
\includegraphics[width=0.24\textwidth]{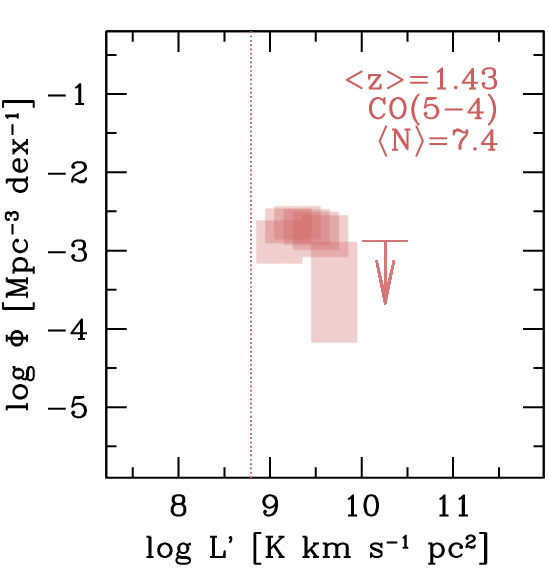}
\includegraphics[width=0.24\textwidth]{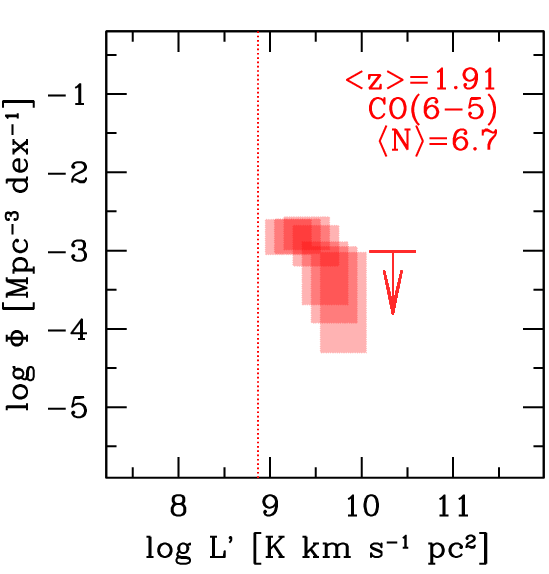}\\
\includegraphics[width=0.24\textwidth]{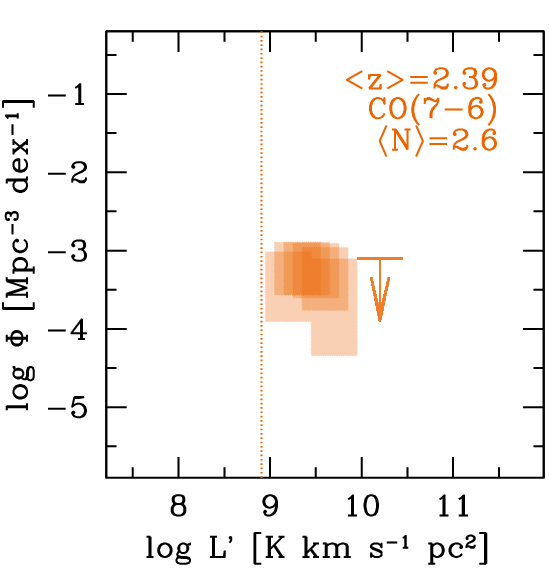}
\includegraphics[width=0.24\textwidth]{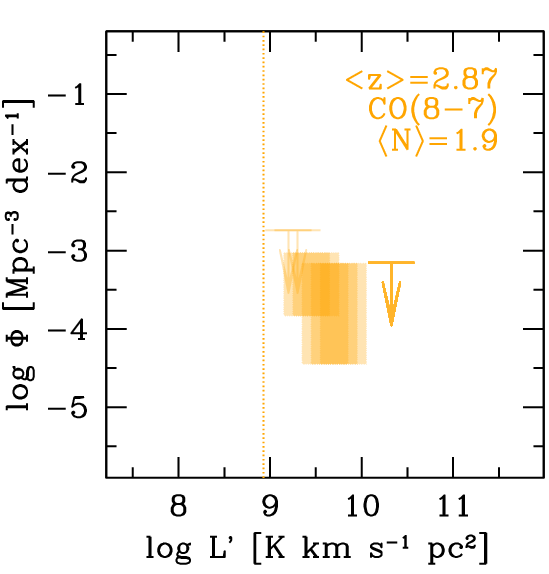}
\includegraphics[width=0.24\textwidth]{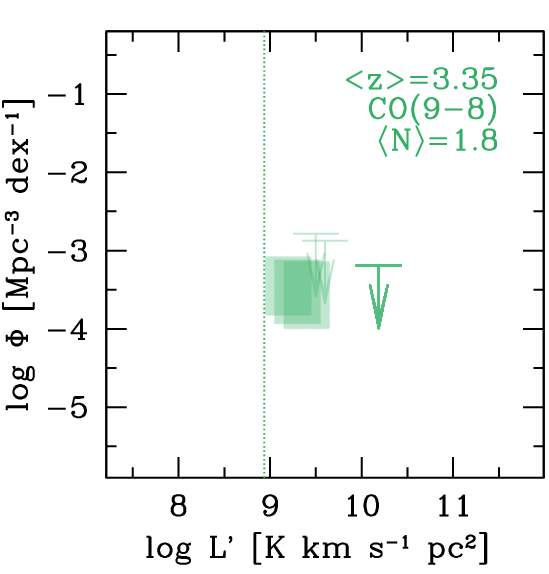}
\includegraphics[width=0.24\textwidth]{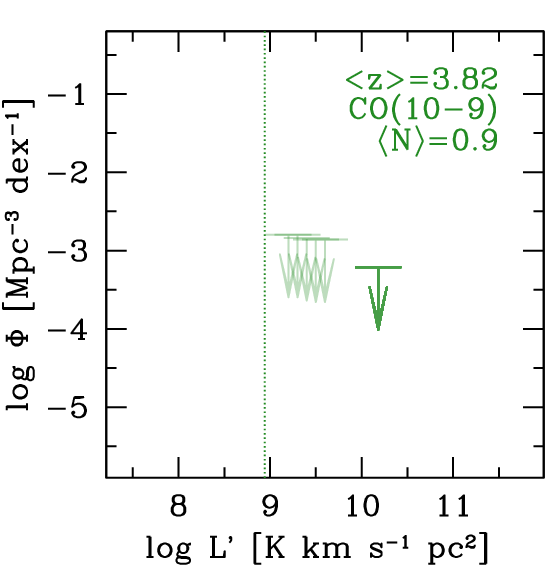}\\
\includegraphics[width=0.24\textwidth]{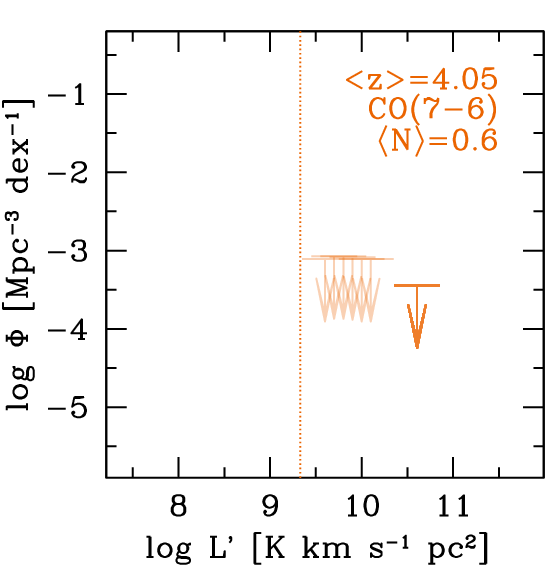}
\includegraphics[width=0.24\textwidth]{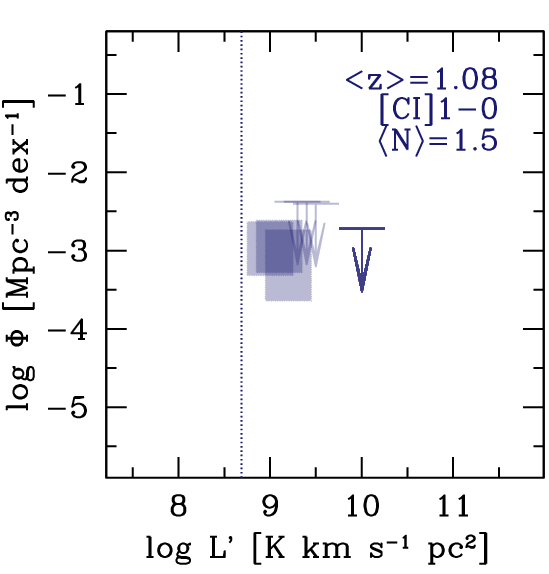}
\includegraphics[width=0.24\textwidth]{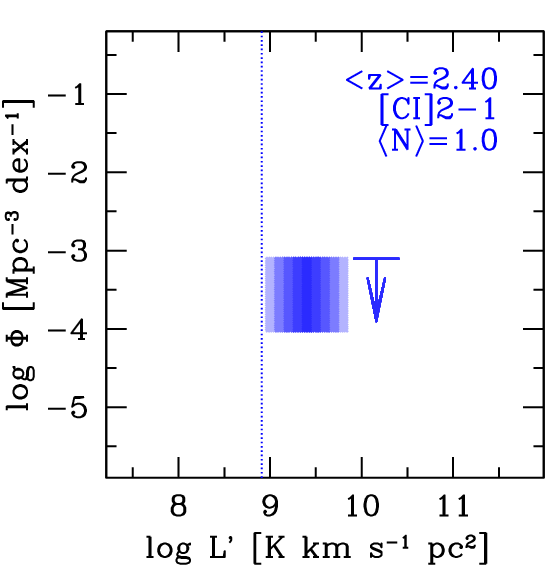}
\includegraphics[width=0.24\textwidth]{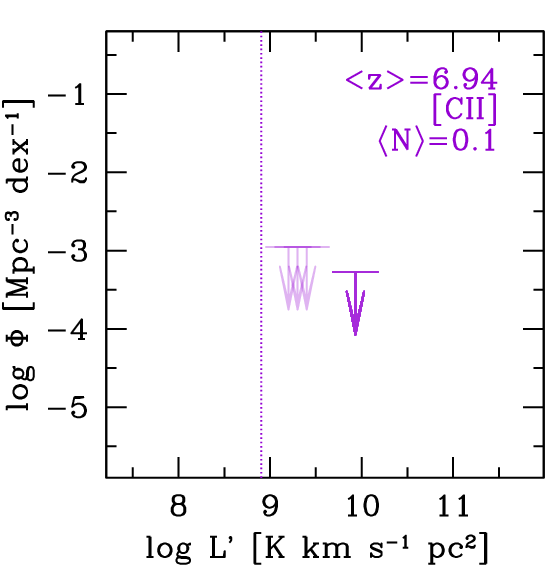}\\
\end{center}
\caption{Constraints on the CO, \Ci{}, and \Cii{} luminosity functions from ASPECS. The vertical extent of the boxes shows the average $\pm$1-$\sigma$ range in each 0.5\,dex bin. For each transition, we report the volume--averaged redshift, and the average number of line candidates used in the various LF realizations. Bins with an average of $>1$ line candidate entry per realization are shown as boxes, while arrows mark the corresponding 3-$\sigma$ limits for all of the other bins. The vertical bars show the  formal 5-$\sigma$ line luminosity limit (see Table~\ref{tab_zbin}).}
\label{fig_co_lf}
\end{figure*}

\subsection{Luminosity functions and $\rho_{\rm H2}$}\label{sec_masses}

In the construction of the CO luminosity functions, we follow the approach adopted in \citet{decarli19}. Namely, we create 5000 realizations of the luminosity functions, folding in all of the uncertainties: formal flux measurement errors from the Gaussian fit, the uncertainties in the line identification (and the implications in terms of luminosity distance), the probability of a line to be spurious (as quantified via the fidelity), etc. In each realization, we keep only a subset of line candidates, based on their fidelity: We extract a number between 0 and 1 from a uniform distribution, and if the value is smaller than the line fidelity, we keep the line candidate in that realization. The resulting catalogs of lines are binned in luminosity, using 0.5\,dex bins. Poissonian uncertainties are estimated for each bin, following \citet{gehrels86}. The number of entries and its uncertainties are then scaled to account for completeness and divided by the effective volume of the survey. Following \citet{riechers19} and \citet{decarli19}, we create five versions of the luminosity functions, shifted by 0.1\,dex one from the other, in order to expose the intra-bin variations despite the modest statistics in each bin. The luminosity functions (and their uncertainties) thus obtained are then averaged among all the realizations.

Fig.~\ref{fig_co_lf} shows the resulting luminosity functions for each transition considered in this study: CO $J_{\rm up}=1$ to 4, and \Ci$_{1-0}$ from 3\,mm, and CO $J_{\rm up}=3$ to 10, \Ci$_{1-0}$, \Ci$_{2-1}$, as well as \Cii{}. Tabulated values are reported in appendix \ref{sec_tables}. We limit our analysis to line candidates brighter than the formal 5-$\sigma$ limit (see Fig.~\ref{fig_limits} and Table~\ref{tab_zbin}), and we only plot bins that are fully accommodated above this luminosity threshold and have an average of at least one entry throughout the realizations.

Finally, we convert the CO(1-0) -- CO(4-3) line luminosities observed in either ASPECS band into H$_2$ masses as described in the previous subsection, we sum over the line candidates used in each realization of the luminosity function, and thus we infer the total molecular gas per cosmological volume, $\rho_{\rm H2}$ (see Tab.~\ref{tab_rhoH2_z}). We remark that in the estimate of $\rho_{\rm H2}$, we do not extrapolate the LFs outside the observed line luminosity ranges, but rather sum over the individual detections (corrected for fidelity and completeness).

\begin{table}
\caption{\rm The cosmic molecular gas density (mass of molecular gas in galaxies per cosmological volume) as constrained by ASPECS.} \label{tab_rhoH2_z}
\begin{center}
\begin{tabular}{ccc}
\hline
 Redshift               & $\rho_{\rm H2}$, 1$\sigma$ & $\rho_{\rm H2}$, 2$\sigma$  \\
                        & [$10^7$ \Msun{}\,Mpc$^{-3}$] & [$10^7$ \Msun{}\,Mpc$^{-3}$]  \\
 (1)    &     (2)       & (3)         \\
\hline
\multicolumn{3}{l}{\it new from ASPECS LP 1.2\,mm}      \\
\multicolumn{1}{l}{\it from CO}       & &               \\
      0.271---0.631   &  0.572---2.148  & 0.286---3.181 \\
      0.695---1.174   &  2.772---7.371  & 1.652---10.02 \\
\multicolumn{1}{l}{\it from \Ci}      & &               \\
      0.809---1.321   &  0.210---1.397  & 0.078---2.240 \\
      1.975---2.816   &  0.150---2.882  & 0.020---4.977 \\
\hline
\multicolumn{3}{l}{\it updated from ASPECS LP 3\,mm, from CO}    \\
      0.003---0.369   & 0.015---0.281   & 0.002---0.485 \\
      1.006---1.738   & 4.053---7.489   & 2.953---9.462 \\
      2.008---3.107   & 1.844---4.438   & 1.164---6.007 \\
      3.011---4.475   & 1.686---3.289   & 1.193---4.220 \\
\hline
\end{tabular}
\end{center}
\end{table}

\section{Discussion}\label{sec_discussion}

\begin{figure*}
\begin{center}
\includegraphics[width=0.24\textwidth]{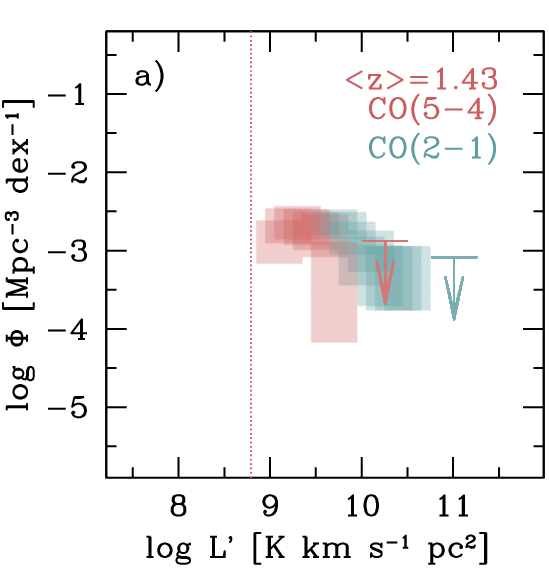}
\includegraphics[width=0.24\textwidth]{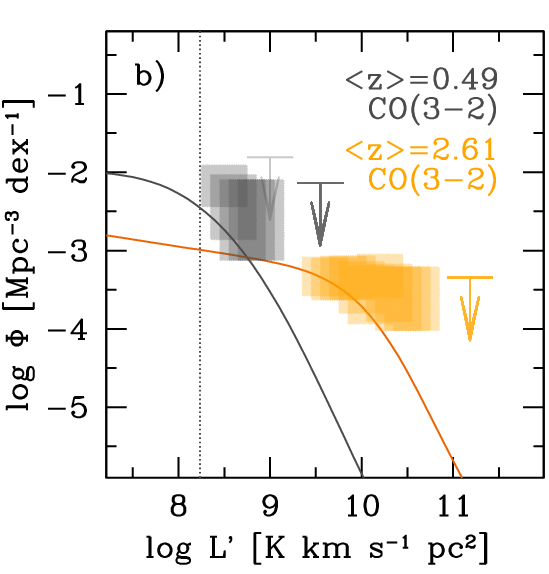}
\includegraphics[width=0.24\textwidth]{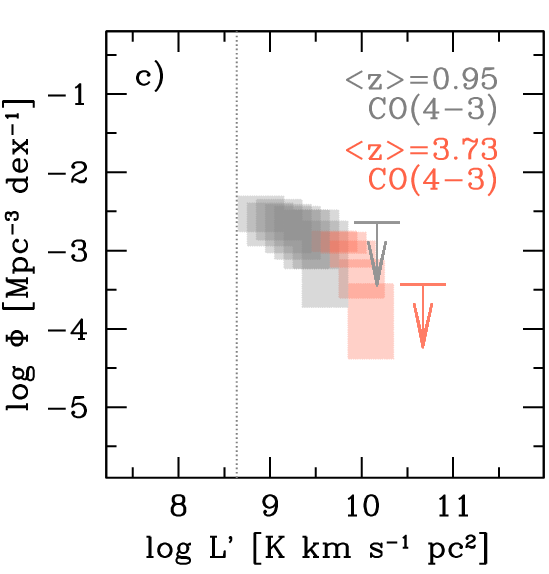}
\includegraphics[width=0.24\textwidth]{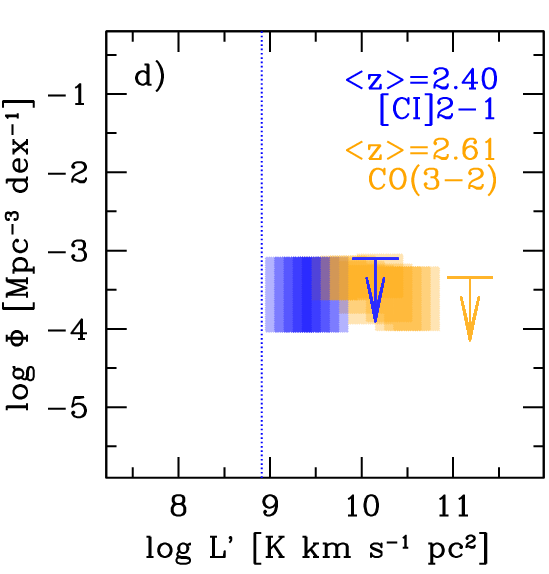}\\
\end{center}
\caption{Comparison between CO and \Ci$_{2-1}$ luminosity functions. a) The CO(2-1) (grey blue) and CO(5-4) (dark red) luminosity functions in the common redshift range around $\langle z \rangle$=$1.43$. The CO(2-1) luminosity function appears in systematic excess with respect to the CO(5-4), hinting at generally sub-thermalized conditions ($r_{52}<0.3$, see text for details). b) The CO(3-2) luminosity functions observed at 3\,mm and 1.2\,mm at $\langle z\rangle$=2.61 and $\langle z\rangle$=0.49, respectively. For comparison, the empirical predictions of the CO(3-2) luminosity functions based on {\it Herschel} IR luminosity functions by \citet{vallini16} are shown in grey ($z\sim 0$) and orange ($z\sim 2$) lines. The CO(3-2) LF appears to evolve from $z\sim 2.6$ to the present age. ASPECS data point to an evolution in the CO(3-2) luminosity function consistent with the empirical predictions, although the difference in the sampled luminosity ranges in the two redshift bins limits the robustness of this finding. c) Similar to the previous panel, but for the CO(4-3) luminosity functions observed at $z\sim0.95$ at 1.2\,mm and at $z\sim 3.7$ at 3\,mm. d) Comparison between the \Ci$_{2-1}$ and CO(3-2)  luminosity functions at $z\sim2.5$. We find an offset of $\gsim 0.5$\,dex between the two luminosity functions, broadly in agreement with similar ratios between the two line luminosities reported in the literature from studies of individual sources (see Sec.~\ref{sec_carbonLF}). }
\label{fig_lf_3mm_1mm}
\end{figure*}

\subsection{CO luminosity functions}

Fig.~\ref{fig_co_lf} shows the constraints on the luminosity functions for all the transitions covered in our analysis. Multiple lines are identified for all the mid--$J$ CO transitions (up to $J_{\rm up}$=7). The CO(8-7) line is securely detected only in one case in the entire ASPECS volume. None of the higher--$J$ CO lines is significantly detected individually, thus only low--fidelity candidates enter the luminosity function analysis for these transitions. Since our line luminosity limit (in units of $L'$) is rather flat with redshift at $z>1$ (see Fig.~\ref{fig_limits}), this result per se can be attributed to sub-thermalized conditions in the ISM of typical galaxies at least at $J_{\rm up}\gsim 7$ or a drop in the gas masses or metallicities of galaxies at $z>1$. In the following section we further explore these scenarios.

\subsubsection{Same redshift, different CO transition}

Fig.~\ref{fig_lf_3mm_1mm} compares the CO luminosity function constraints from the two bands of ASPECS. At $\langle z \rangle\approx1.43$, the ASPECS frequency coverage is such that we observe the CO(2-1) transition at 3\,mm and the CO(5-4) transition at 1.2\,mm. The inferred CO luminosity functions show an offset of about 0.5 dex in luminosity for a fixed number density. This immediately implies sub-thermalized conditions of the molecular ISM in the targeted galaxies ($r_{52}\lsim 0.3$, consistent with the value of $r_{52}\approx 0.16$ derived by \citealt{boogaard20}).

\subsubsection{Same CO transition, different redshifts}

The ASPECS frequency coverage also allows us to trace the same line transition, CO(3-2), both at $\langle z \rangle\approx 0.49$ at 1.2\,mm, and at $\langle z \rangle\approx 2.61$ at 3\,mm. Because of the $\sim 16.2\times$ smaller volume and $\sim 7.7\times$ lower luminosity distance, we sample different ranges of the CO(3-2) luminosity function in the two redshift bins, with the low--redshift data mostly constraining the $L'<10^9$\,\Kkmspc{} regime and the high--redshift data pinning down the bright end at $L'>2\times 10^9$\,\Kkmspc{}. However, the difference in number density throughout the observed range strongly points towards an evolution of the CO(3-2) luminosity function between $z\sim 2.6$ and $z\sim 0.5$. This is even clearer once we compare the observed CO LFs with the empirical predictions based on the {\it Herschel} IR LFs from \citet{vallini16} shown in Fig.~\ref{fig_lf_3mm_1mm}. The {\it Herschel} IR LFs were scaled via an empirical relation of the form: $\log L'$/(\Kkmspc)$ = 0.54 + 0.81 \log L_{\rm IR}/$\Lsun{} \citep{sargent14}. The observed CO LF at $z\sim 0.5$ appears to sample just above the expected knee of the CO LFs. The observed CO(3-2) LF at $z\sim 2.6$ is in good agreement with the prediction for $z\sim2$ around the expected knee, and it lies $>$2 dex higher (in terms of number density) than the low--$z$ predictions for $L'\sim 10^{10}$\,\Kkmspc. This result provides further, direct support to an evolution in the CO LFs, and therefore in gas content of galaxies, in this case irrespective of uncertainties in the CO excitation. We also show the comparison between the CO(4-3) LFs observed at $z\sim 0.95$ at 1.2\,mm and $z\sim3.7$ at 3\,mm. A similar LF evolution might also be present in CO(4-3), but the available data do not allow us to exclude a non-evolving scenario.

\subsection{\Ci{} and \Cii{} luminosity functions}\label{sec_carbonLF}

In Fig.~\ref{fig_co_lf}, we also show the observed constraints on the \Ci$_{1-0}$ LF at $\langle z \rangle$=1.08, on the \Ci$_{2-1}$ LF at $\langle z \rangle$=2.40, and on the \Cii{} LF at $\langle z \rangle$=6.94 from ASPECS 1.2\,mm. 
With the exception of the strong \Ci{}$_{2-1}$ detection associated with the galaxy ASPECS LP 1mm.C01 \citep{boogaard20,aravena20}, only relatively low fidelity candidates are consistent with being \Ci{} or \Cii{} transitions. We further explore ASPECS contraints on the \Cii{} LF in \citet{uzgil20}.

Fig.~\ref{fig_lf_3mm_1mm} shows the comparison between the \Ci$_{2-1}$ LF from our 1.2\,mm cube, and the CO(3-2) LF from the ASPECS LP 3\,mm. The two LFs probe roughly the same redshift range, so the comparison of the two LFs yields an insight on the average physical conditions in the ISM of the detected galaxies. We find a global shift of $\gsim 0.5$\,dex between the two LFs, which is roughly consistent with the median ratio of $0.69\pm0.16$\,dex for \Ci$_{2-1}$/CO(3-2) reported in targeted observations of SMGs and quasar host galaxies at $z$=2--6 in \citet{walter11}. For comparison, \citet{jiao17} find a ratio of 0.9\,dex in local ULIRGs. 

We refer to \citet{boogaard20} for a more detailed discussion of the astrophysical implication of the observed CO to \Ci{} line ratios, and to \citet{uzgil20} for a further exploration of the upper limits on the \Cii{} LF.

\begin{figure*}
\begin{center}
\includegraphics[width=0.8\textwidth]{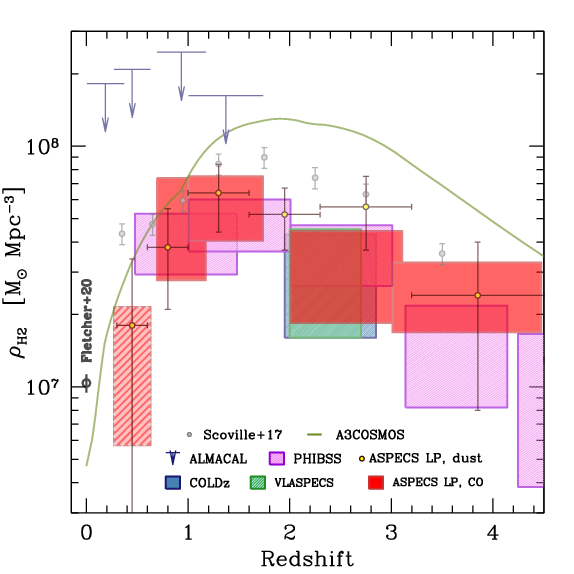}
\end{center}
\caption{The evolution of the cosmic molecular gas density, $\rho_{\rm H2}$($z$), from ASPECS LP compared to similar studies in the literature: CO--based measurements from VLASPECS \citep{riechers20}, COLDz \citep{riechers19}, PHIBSS fields \citep{lenkic20}, ALMACAL \citep{klitsch19}; and dust--based measurements from ASPECS \citep{magnelli20}, A3COSMOS \citep{liu19}, and from \citet{scoville17} (see footnote 2). The $\rho_{\rm H2}$($z$=0) measurement by \citet{fletcher20} is also shown for reference. All of the uncertainties are shown at 1-$\sigma$ significance. The ASPECS LP constraints at $z\lsim 0.5$ are shaded to highlight the non-negligible impact of cosmic variance at these redshifts. The available datasets all point towards a steep decrease in $\rho_{\rm H2}$ from cosmic noon to the local universe preceded by a smooth increase from higher redshift. Different surveys targeting different regions of the sky appear to find the same trend, implying that cosmic variance does not dominate the results (see Appendix \ref{sec_cvariance}).}
\label{fig_rhoH2_comp}
\end{figure*}

\subsection{$\rho_{\rm H2}$ vs redshift}

We use the combined ASPECS data to infer the cosmic--averaged molecular gas density of galaxies, $\rho_{\rm H2}$, as a function of cosmic time (see Sec.~\ref{sec_masses}). Compared to previous incarnations of our analysis \citep[e.g.,][]{walter14,decarli16a,decarli19}, we here adopt the updated constraints on the CO excitation from \citet{boogaard20}, which also includes the VLASPECS results \citep{riechers20}. Our analysis yields a nearly continuous sampling of $\rho_{\rm H2}$($z$) from $z\approx0$ to $z\sim 4.5$ in a self-consistent manner. 
The ASPECS data show a smooth increase of $\rho_{\rm H2}$($z$) from early cosmic time up to $z\sim 1.5$, followed by a $\sim 6\times$ decline to the present day (see Fig.~\ref{fig_rhoH2_comp} and Table~\ref{tab_rhoH2_z}). The new excitation correction \citep{boogaard20} brings the $\rho_{\rm H2}$($z$) constraints from CO into excellent agreement with our dust-based measurements from ASPECS \citep{magnelli20}. The $\rho_{\rm H2}$ constraints at $z\lsim 0.5$ from ASPECS are rather loose, as a result of the small volume probed (see Appendix~\ref{sec_cvariance}).

We note that the results shown in Fig.~\ref{fig_rhoH2_comp} are based on a constant $\alpha_{\rm CO}$ or gas--to--dust ratio. The arguments presented in Sec.~\ref{sec_measures} for a Galactic value may not be valid at $z\gsim3$, where we lack direct constraints on the metallicity of typical CO-- and dust--emitting galaxies. A lower metallicity would imply a higher $\alpha_{\rm CO}$ and gas--to--dust ratio, yielding to higher $\rho_{\rm H2}$ estimates.

We also derive \Ci{}--based estimates of $\rho_{\rm H2}$($z$) (see Table~\ref{tab_rhoH2_z}). The two \Ci{}--based estimates at $z\sim1$ and $z\sim 2.5$ appear lower by a factor $\sim 5\times$ and $\sim2\times$, respectively, compared to the corresponding CO--based estimates. This discrepancy is likely due to sensitivity limitations, and highlights the challenge of using \Ci{} as molecular gas tracer of the bulk of the galaxy population at high redshift \citep[for dedicated \Ci{} studies in main sequence galaxies at high redshift, see, e.g.,][]{valentino18,valentino20}.

In Fig.~\ref{fig_rhoH2_comp} we place the ASPECS measures of $\rho_{\rm H2}$($z$) in the context of similar investigations in the literature. Our new measurements, listed in Table~\ref{tab_rhoH2_z}, improve and expand on the results from previous molecular scans using the Plateau de Bure Interferometer \citep{walter14}, the VLA \citep{riechers19}, and ALMA \citep{decarli16a,decarli19}, as well as the constraints from field sources in the PHIBSS data \citep{lenkic20}, and from calibrator fields in the ALMACAL survey \citep{klitsch19}. Our comparison also includes dust-based $\rho_{\rm H2}$($z$) measurements from \citet{scoville17}, \citet{liu19}, and from ASPECS \citep{magnelli20}. Overall, the molecular gas constraints from volume--limited surveys agree within the uncertainties over $\sim 90$\% of the cosmic history. The general agreement in these results, based on different fields, suggests that the impact of cosmic variance and of systematics is modest. In Appendix \ref{sec_cvariance} we quantitatively assess its role within our dataset. The studies by \citet{scoville17} and \citet{liu19} find a qualitatively similar evolution of $\rho_{\rm H2}$($z$), although with different normalizations. These $\rho_{\rm H2}$ estimates rely on different assumptions of stellar mass functions, functional form of the main sequence, gas fractions, internal calibrations, and integration limits. Homogenizing these is beyond the scope of the present work, therefore we here only show their `bona fide' estimates as published.

The observed evolution of $\rho_{\rm H2}$ appears to mimic the history of the cosmic star formation rate density, $\rho_{\rm SFR}$ \citep[see, e.g.,][]{madau14}. The ratio between $\rho_{\rm H2}$ and $\rho_{\rm SFR}$ results in a volume--average of the ``depletion time'' $\langle t_{\rm dep} \rangle$, i.e., the timescale required for galaxies to deplete their reservoirs of molecular gas, if star formation continues at the current rate, and there is no further gas accretion or outflows. Our results hint to a relatively constant $\langle t_{\rm dep} \rangle$. In \citet{walter20} we explore the astrophysical implications of this result in the context of galaxy evolution.

\section{Conclusions}\label{sec_conclusions}

We present the ultimate CO luminosity functions from the ASPECS large program, and the resulting constraints for the cosmic evolution of the molecular gas density. The main conclusions of this study of the molecular and atomic line emission in ASPECS LP are as follows.

\begin{itemize}
\item[{\em i-}] The line flux distributions due to various CO and neutral/ionized Carbon lines in our analysis show that roughly 80\% of the line flux at 3\,mm is associated with CO(2-1) or CO(3-2) at the age of cosmic noon, and $60$\% of the line flux at 1.2\,mm is due to intermediate--$J$ CO transition ($J_{\rm up}$=3--6) at $z\lsim2$. Higher--$J$ CO transitions are negligible at 3\,mm but account for 25\% of the total line flux at 1.2\,mm. Neutral Carbon contributes to $\sim12$\% of the integrated line flux at 1.2\,mm. Finally, singly-ionized Carbon \Cii{} at $6\lsim z \lsim 8$ accounts for $< 1$\% of the line flux at 1.2\,mm. This result poses a major challenge for intensity mapping experiments targeting \Cii{} at the end of the epoch of reionization, as the expected line foreground is two orders of magnitudes stronger (in terms of total flux in lines) than the \Cii{} signal.
\item[{\em ii-}] The CO luminosity functions probed at 1.2\,mm evolve as a function of redshift, with a decrease in the number density at a given line luminosity (in units of $L'$). This implies substantially sub-thermal excitation in galaxies throughout the last $\sim$10 Gyr of cosmic history.
\item[{\em iii-}] The direct comparison between the luminosity functions for the same CO transition seen in the 1.2\,mm and 3\,mm cubes of the ASPECS LP reinforces the idea that the typical galaxy at $z\approx 1.43$ shows sub-thermalized molecular gas emission, and that there is a significant evolution in the luminosity function for CO(2-1) takes place between $z\sim2.8$ and $z\sim 0.5$ irrespective of any CO excitation assumption. A comparison between the \Ci{}$_{2-1}$ and CO(3-2) luminosity functions in the redshift range $z\sim 2.5$ suggests that the line ratio is in line with the values reported for IR--bright galaxies in targeted studies.
\item[{\em iv-}] The cosmic density of molecular gas in galaxies, $\rho_{\rm H2}$, smoothly increases from early cosmic time up $z\sim 2-3$, followed by a factor $\sim 6$ drop to the present age. This is in qualitative agreement with the cosmic SFR density, suggesting that the depletion time of galaxies is approximately constant in redshift once averaged over the galaxy population.
\item[{\em v-}] Modeling and the comparison with similar surveys suggest that cosmic variance does not play a dominant role in our estimates of $\rho_{\rm H2}$ at $z\gsim 0.5$.
\end{itemize}

The emerging consensus on the evolution of $\rho_{\rm H2}$ is the result of many hundreds of hours of integration with PdBI/NOEMA, VLA, and ALMA. Using these facilities to significantly expand on the latest campaigns is still possible, but observationally expensive. Future upgrades in the capabilities of available instruments (from the forthcoming completion of NOEMA, to the plans outlined in the ALMA 2030 Roadmap,  \citealt{carpenter20}, and in the next generation VLA white books, \citealt{murphy18}) are required in order to make the next transformational step in this field.

\acknowledgements

FW and MN acknowledge support by the ERC Advanced Grant Cosmic-Gas (740246). Este trabajo cont \'o con el apoyo de CONICYT + PCI + INSTITUTO MAX PLANCK DE ASTRONOMIA MPG190030. TD-S acknowledges support from the CASSACA and CONICYT fund CAS-CONICYT Call 2018. JH acknowledges support of the VIDI research programme with project number 639.042.611, which is (partly) financed by the Netherlands Organisation for Scientific Research (NWO). DR acknowledges support from the National Science Foundation under grant numbers AST-1614213 and AST-1910107 and from the Alexander von Humboldt Foundation through a Humboldt Research Fellowship for Experienced Researchers. HI acknowledges support from JSPS KAKENHI Grant Number JP19K23462. IRS acknowledges support from STFC (ST/T000244/1).

\facility{ALMA} data: 2016.1.00324.L. ALMA is a partnership of ESO (representing its member states), NSF (USA) and NINS (Japan), together with NRC (Canada), NSC and ASIAA (Taiwan), and KASI (Republic of Korea), in cooperation with the Republic of Chile. The Joint ALMA Observatory is operated by ESO, AUI/NRAO and NAOJ.

\appendix

\section{Tabulated luminosity functions}\label{sec_tables}

Tables \ref{tab_co_lf} and \ref{tab_carbon_lf} list the ASPECS constraints on the luminosity functions of CO, \Ci{} and \Cii{}.

\begin{table*}
\caption{\rm Luminosity functions of the observed CO transitions. (Columns: 1, 3, 5) Luminosity bin center; each bin is 0.5\,dex wide. (Columns: 2, 4, 6) minimum and maximum values of the luminosity function confidence levels at 1-$\sigma$, or 3-$\sigma$ upper limits on the luminosity functions.} \label{tab_co_lf}
\vspace{-0.5cm}
 \begin{center}
 \begin{tabular}{cc|cc|cc}
 \hline
 log $L'$ & log $\Phi$ &  log $L'$ & log $\Phi$ &  log $L'$ & log $\Phi$  \\
  {}[K\,km\,s$^{-1}$\,pc$^2$] & [dex$^{-1}$\,Mpc$^{-3}$] & {}[K\,km\,s$^{-1}$\,pc$^2$] & [dex$^{-1}$\,Mpc$^{-3}$] & {}[K\,km\,s$^{-1}$\,pc$^2$] & [dex$^{-1}$\,Mpc$^{-3}$] \\
  (1)  & (2)          & (3)           & (4)   & (5)           & (6)   \\
\hline
 \multicolumn{2}{c}{CO(1-0), 3\,mm} & \multicolumn{2}{c}{CO(3-2), 1.2\,mm} & \multicolumn{2}{c}{CO(7-6), 1.2\,mm}\\
 8.5  & $<$ $-$1.38      &  8.5  & $-$2.44 $-$1.91  &    9.2  & $-$3.91 $-$3.02 \\
 8.6  & $<$ $-$1.38      &  8.6  & $-$2.86 $-$2.03  &    9.3  & $-$3.57 $-$2.89 \\
 8.7  & $<$ $-$1.38      &  8.7  & $-$3.12 $-$2.09  &    9.4  & $-$3.57 $-$2.89 \\
 8.8  & $<$ $-$1.38      &  8.8  & $-$3.12 $-$2.09  &    9.5  & $-$3.60 $-$2.91 \\
 8.9  & $<$ $-$1.38      &  8.9  & $-$3.12 $-$2.09  &    9.6  & $-$3.76 $-$2.96 \\
 9.0  & $<$ $-$1.40      &  9.0  & $<$ $-$1.81      &    9.7  & $-$4.34 $-$3.11 \\
\hline
 \multicolumn{2}{c}{CO(2-1), 3\,mm} & \multicolumn{2}{c}{CO(4-3), 1.2\,mm} & \multicolumn{2}{c}{CO(8-6), 1.2\,mm}\\
 9.5  & $-$2.87 $-$2.53  &  8.9  & $-$2.76 $-$2.30  &    9.2  & $<$ $-$2.74     \\
 9.6  & $-$2.88 $-$2.53  &  9.0  & $-$2.94 $-$2.39  &    9.3  & $<$ $-$2.74     \\
 9.7  & $-$2.80 $-$2.48  &  9.1  & $-$2.87 $-$2.35  &    9.4  & $-$3.83 $-$3.03 \\
 9.8  & $-$2.85 $-$2.51  &  9.2  & $-$3.03 $-$2.41  &    9.5  & $-$3.83 $-$3.03 \\
 9.9  & $-$3.05 $-$2.63  &  9.3  & $-$3.11 $-$2.44  &    9.6  & $-$4.45 $-$3.16 \\
10.0  & $-$3.44 $-$2.83  &  9.4  & $-$3.23 $-$2.48  &    9.7  & $-$4.45 $-$3.16 \\
10.1  & $-$3.27 $-$2.75  &  9.5  & $-$3.23 $-$2.48  &    9.8  & $-$4.45 $-$3.16 \\
10.2  & $-$3.71 $-$2.93  &  9.6  & $-$3.72 $-$2.62  &	      &                 \\
10.3  & $-$3.76 $-$2.95  &	 &		    &	      &                 \\
10.4  & $-$3.76 $-$2.95  &	 &		    &	      &                 \\
10.5  & $-$3.76 $-$2.95  &	 &		    &	      &                 \\
\hline
 \multicolumn{2}{c}{CO(3-2), 3\,mm} & \multicolumn{2}{c}{CO(4-3), 1.2\,mm} & \multicolumn{2}{c}{CO(9-8), 1.2\,mm}\\
 9.6  & $-$3.58 $-$3.08  &  9.1  & $-$3.16 $-$2.62  &    9.2  & $-$3.83 $-$3.08 \\
 9.7  & $-$3.63 $-$3.09  &  9.2  & $-$2.90 $-$2.46  &    9.3  & $-$3.93 $-$3.12 \\
 9.8  & $-$3.63 $-$3.07  &  9.3  & $-$2.86 $-$2.44  &    9.4  & $-$4.00 $-$3.14 \\
 9.9  & $-$3.62 $-$3.07  &  9.4  & $-$2.93 $-$2.47  &    9.5  & $<$ $-$2.78     \\
10.0  & $-$3.80 $-$3.13  &  9.5  & $-$2.99 $-$2.51  &    9.6  & $<$ $-$2.87     \\
10.1  & $-$3.94 $-$3.19  &  9.6  & $-$3.08 $-$2.55  &	      &                 \\
10.2  & $-$3.60 $-$3.04  &  9.7  & $-$4.17 $-$2.89  &	      & 	        \\
10.3  & $-$3.91 $-$3.17  &	 &		    &	      & 	        \\
10.4  & $-$4.02 $-$3.21  &	 &		    &	      & 	        \\
10.5  & $-$4.02 $-$3.21  &	 &		    &	      &  	        \\
10.6  & $-$4.02 $-$3.21  &	 &		    &	      & 	        \\
\hline
 \multicolumn{2}{c}{CO(4-3), 3\,mm} & \multicolumn{2}{c}{CO(5-4), 1.2\,mm} & \multicolumn{2}{c}{CO(10-9), 1.2\,mm}\\
 9.7  & $-$3.01 $-$2.75  &  9.2  & $-$3.05 $-$2.60  &    9.2  & $<$ $-$2.80   \\
 9.8  & $-$3.03 $-$2.77  &  9.3  & $-$3.04 $-$2.60  &    9.3  & $<$ $-$2.80   \\
 9.9  & $-$3.21 $-$2.88  &  9.4  & $-$2.99 $-$2.57  &    9.4  & $<$ $-$2.84   \\
10.0  & $-$3.61 $-$3.12  &  9.5  & $-$3.20 $-$2.68  &    9.5  & $<$ $-$2.86   \\
10.1  & $-$4.38 $-$3.42  &  9.6  & $-$3.69 $-$2.89  &    9.6  & $<$ $-$2.86   \\
10.2  &  $<$ $-$3.09     &  9.7  & $-$3.92 $-$2.95  &	      &               \\
      &                  &  9.8  & $-$4.31 $-$3.02  &	      &               \\
\hline
\end{tabular}
\end{center}
\end{table*}

\begin{table}
\caption{\rm Luminosity functions of the observed \Ci{} and \Cii{} lines. (Columns: 1, 3) Luminosity bin center; each bin is 0.5\,dex wide. (Columns: 2, 4) minimum and maximum values of the luminosity function confidence levels at 1-$\sigma$, or 3-$\sigma$ upper limits on the luminosity functions.} \label{tab_carbon_lf}
\vspace{-0.5cm}
 \begin{center}
 \begin{tabular}{cc|cc}
 \hline
 log $L'$ & log $\Phi$  &  log $L'$ & log $\Phi$  \\
  (1)  & (2)           &  (3)  & (4)            \\
\hline
 \multicolumn{2}{c}{\Ci{}$_{1-0}$, 3\,mm}  & \multicolumn{2}{c}{\Ci{}$_{2-1}$, 1.2\,mm} \\
 9.6  & $<$ $-$3.10      &	9.2  & $-$4.04 $-$3.08  \\ 
 9.7  & $<$ $-$3.07      &	9.3  & $-$4.04 $-$3.08  \\ 
 9.8  & $<$ $-$3.07      &	9.4  & $-$4.04 $-$3.08  \\ 
 9.9  & $<$ $-$3.08      &	9.5  & $-$4.04 $-$3.08  \\ 
10.0  & $<$ $-$3.11      &	9.6  & $-$4.04 $-$3.08  \\ 
10.1  & $<$ $-$3.11      &           &                  \\
\hline
 \multicolumn{2}{c}{\Ci{}$_{1-0}$, 1.2\,mm}  & \multicolumn{2}{c}{\Cii{}, 1.2\,mm} \\
 9.0  & $-$3.32 $-$2.63  &      9.1  & $<$ $-$2.95      \\
 9.1  & $-$3.28 $-$2.61  &	9.2  & $<$ $-$2.95      \\
 9.2  & $-$3.64 $-$2.73  &	9.3  & $<$ $-$2.95      \\
 9.3  & $<$ $-$2.38      &	9.4  & $<$ $-$2.95      \\
 9.4  & $<$ $-$2.38      &	9.5  & $<$ $-$2.97      \\
\hline 
\end{tabular}
\end{center}
\end{table}

\section{Cosmic variance}\label{sec_cvariance}

A critical limitation of pencil--beam surveys such as ASPECS is the impact of cosmic variance. Noticeably, a large fraction of the galaxies detected in CO(2-1) emission in ASPECS LP 3\,mm belongs to a large overdensity at $z\approx 1.09$ \citep[see][]{boogaard19}. Here we quantify how the clustering of sources impact our results. The expected number of galaxies in a volume--limited survey is:
\begin{equation}\label{eq_Nexp}
N = \int_{V_1}\,\int_{V_2}\,(1+\xi)\,n_1 n_2 \, dV_1\,dV_2
\end{equation}
where $n_i$ is the number density of galaxies, obtained by integrating the luminosity (or mass) function of galaxies down to the detection threshold of the survey, $V_i$ is the survey volume, and $\xi$ is the 3D 2-points correlation function, which accounts for the excess of galaxy counts compared to the average field due to galaxy clustering. In the linear clustering regime, $\xi$ is often modeled as a power-law: $\xi(r)=(r/r_0)^{-\gamma}$. The variance on the expected numbers, Var[$N$], is usually referred to as cosmic variance. It comprizes of a Poissonian term, and a term due to the variations in the number counts due to clustering:
\begin{equation}\label{eq_cvar}
\sigma_v^2=\frac{\langle N^2\rangle-\langle N\rangle^2-\langle N\rangle}{\langle N \rangle^2} = \frac{1}{V^2} \int_{V_1}\,\int_{V_2}\, \xi \, dV_1\,dV_2
\end{equation}
As discussed in \citet{decarli16a} and \citet{decarli19}, the Poissonian uncertainties are accounted for in the construction of the CO luminosity functions and in our estimates of $\rho_{\rm H2}$. The clustering term in Eq.~\ref{eq_cvar} implies that, even in presence of large source counts, field--to--field variations are expected due to large--scale structures and clustering. This might introduce a systematic bias in the estimates of LFs based on datasets centered on pre-selected targets \citep[see also][]{loiacono20}. Here we quantify how our results depend on the choice of the targeted region. 

Directly solving the integral in Eq.~\ref{eq_cvar} would require assumptions on the clustering of CO--bright sources, for which no direct observational constraint is available yet. An alternative and commonly--adopted approach is to rely on theoretical models of galaxy formation to create multiple realizations of galaxy populations in various volume samplings. Cosmic variance is then directly computed using the actual variations of $N$. Here we follow the latter method by capitalizing on data--driven simulations presented in \citet{popping20}. From these simulations we create 100 samplings of the simulated box with a geometry matched to the ASPECS survey volume. We then apply different cuts on the galaxy samples to mimic the selection criteria of ASPECS (see below). Finally, we compute the average and variance in the number of selected galaxies from all the realizations. The variance is a combination of the intrinsic scatter due to the cosmic structures within the simulation, and of Poissonian scattering. The contribution of the latter is directly computed following \citet{gehrels86}, thus we can infer the impact of large-scale structures in the count rates used in our LFs. 

\begin{figure}
\begin{center}
\includegraphics[width=0.49\textwidth]{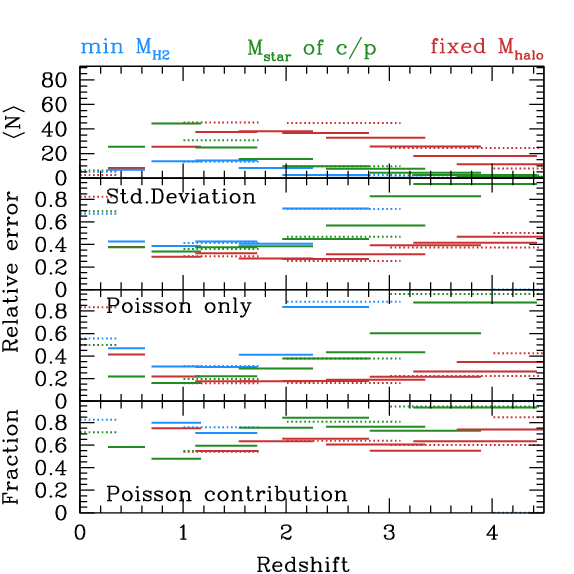}\\
\end{center}
\caption{Impact of cosmic variance on the expected number counts of galaxies in our survey, based on the models presented in \citet{popping20}, as a function of redshift. Blue, green, and red lines show galaxies selected based on the predicted CO luminosity (via the simulated H$_2$ mass), on the stellar mass of their optical/near-infrared counterparts, and on the halo mass, respectively (see text for details). The top panel shows the average number of galaxies in each redshift bin probed with ASPECS LP 1.2\,mm (solid lines) and 3\,mm (dotted lines). The second panel shows the root square of the total cosmic variance. The third panel shows the Poissonian term alone. Finally, the bottom panel shows the fraction of the standard deviation that is due to Poisson. We find that the Poisson contribution dominates the cosmic variance ($>50$\% of the standard deviation) at any redshift, irrespective of the selection function. This implies that the impact of clustering (i.e., the non-Poissonian component of the cosmic variance) is small, and often negligible in ASPECS.}
\label{fig_cvariance}
\end{figure}

The results of this analysis are presented in Fig.~\ref{fig_cvariance}, where we show the average number of galaxies, the standard deviation (i.e., the squared root of the total variance in the number of galaxies), the Poissonian fluctuations, and the fraction of the uncertainties that is attributed to Poissonian fluctuations. Concerning the selection function, for a given transition, ASPECS applies a selection based on the line flux. As this is not trivially derived in models \citep[see extensive discussions in, e.g.,][]{lagos11,popping14,popping19}, here we opt for three different approaches: First, we apply a simple, redshift independent cut in the dark matter halo mass, $M_{\rm halo}>10^{11.5}$\,\Msun. Then we consider a cut based on the minimum stellar mass of detected optical/near-infrared counterparts as a function of redshift. The threshold is $M_{\rm star}>10^{9.0}$\,\Msun{} at $z\approx0.5$, $M_{\rm star}>10^{9.7}$\,\Msun{} at $z\approx1.4$, $M_{\rm star}>10^{10.1}$\,\Msun{} at $z\approx2.4$, $M_{\rm star}>10^{10.3}$\,\Msun{} at $z\approx3.5$, at $M_{\rm star}>10^{10.5}$\,\Msun{} at $z\approx4.5$. Finally, we consider a cut based on the CO 5-$\sigma$ luminosity thresholds shown in Fig.~\ref{fig_limits}, using the predicted CO luminosity in models, based on the simulated H$_2$ mass, under the same $r_{J1}$ and $\alpha_{\rm CO}$ assumptions as used elsewhere in this work (see section \ref{sec_measures}). We find that the number of galaxies we expect to detect is $<15$ in each redshift bin for the CO luminosity cut, while the stellar mass cut at the halo mass cut yield larger numbers of expected galaxies (up to $\sim 50$ around cosmic noon). However, even in these cases, Poissonian uncertainties appear to dominate the total error budget, i.e., the Poisson contribution accounts for $>50$\% of the standard deviation in the number of galaxies at any redshift, irrespective of the selection function. Variance purely due to the large--scale structure in the universe (second panel from the top in Fig.~\ref{fig_cvariance}) plays a significant role only at $z\lsim 0.5$ (in all cases) and $z\gsim 3$--4 (depending on the adopted the adopted selection cut. The overall low impact of cosmic variance is likely to be attributed to the peculiar pencil beam geometry of the survey, with the line--of--sight dimension stretching over $\sim 1000$\,Mpc in most redshift bins. The Poissonian fluctuations are already accounted for in the LFs and estimates of $\rho_{\rm H2}$($z$). The remainder term, due to the clustering of sources, is small in the redshift range of interest, its actual value strongly depends on the (unknown) reliability of our forward--modeling of the selection function. Therefore, we opt not to include this further term into our estimates of the uncertainties. In support to the negligible contribution of cosmic variance, \citet{magnelli20} and \citet{bouwens20} find an excellent match between the stellar mass functions and cosmic SFR density in the ASPECS footprint and the ones inferred in the literature from much wider regions in different (physically disconnected) fields at any $z\gsim 0.5$.

\section{Identification of line candidates without near-infrared counterparts}\label{sec_probs}

Here we explore how our treatment of the line candidates without a counterpart affects our results on the molecular gas content in the universe, $\rho_{\rm H2}$($z$). The expected number of lines from a specific transition is given by the integral over the corresponding above the luminosity limit set in Fig.~\ref{fig_limits}, scaled for the cosmological volume sampled in such transition. In addition, the lack of a counterpart in our multi-wavelength catalog implies an additional, unknown selection function that favors high-redshift scenarios. The modest information content in the data concerning the actual redshift of line candidates without a counterpart implies that the posterior distributions might be affected by our prior assumptions. Hence, we test how different options affect our final results.

Beside our fiducial approach described in Sec.~\ref{sec_masses}, we consider four scenarios: 1) We assume that the probability distribution of line identification is proportional to volume; 2) We assume the same volume--based argument at 3\,mm, and infer expected LFs (and hence, expected number of sources) for 1.2\,mm transitions based on the CO LFs from \citet{saintonge17} (extrapolated to $z\sim0.5$) and from \citet{decarli19} (at $z\gsim0.9$), paired with the large velocity gradient analysis on individual ASPECS sources from \citet{boogaard20}; 3) For both bands we assume that the probability distribution scales according to the flux distribution of line candidates with a counterpart (see Fig.~\ref{fig_flux_distr}); 4) Finally, we restrict our analysis to lines with a 1.2\,mm continuum counterpart \citep[see][]{gonzalezlopez20,aravena20}. These different approaches have their strengths and drawbacks.  The volume--based arguments use the least prior information, but they do not account for the different luminosity limits and for the evolution of the LFs, nor for the intrinsic ratios of line luminosities. The flux--based method has the advantage of resulting in a realistic distribution of the line fluxes for sources without a counterpart, but inherently assumes that the sources with and without a counterpart share a similar redshift and flux distribution, which is unlikely. The forward--modeling method has the advantage of exploiting the information available at 3\,mm and from local studies to constrain the 1.2\,mm LFs, but it relies on extrapolation of observed LFs in different redshift bins, and is partially circular, in that the excitation constraints are based on the same 1.2\,mm data. Finally, limiting the analysis to secure sources provides us with a robust lower limit, but this approach does not fully capitalize on the signal present in the data.

\begin{figure}
\begin{center}
\includegraphics[width=0.49\textwidth]{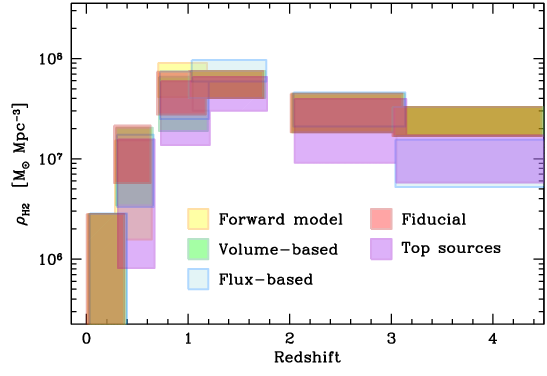}\\
\end{center}
\caption{The evolution of the molecular gas cosmic density in galaxies as a function of redshift, shown under different treatments of the line candidates without a counterpart. 
Red boxes: our bona--fide constraints, as shown in Fig.~\ref{fig_rhoH2_comp}. 
Green boxes: the probability distribution of line identification scales with the sampled volume in each transition, for both ASPECS bands. 
Yellow boxes: the probability distribution of line identification scales as the volume for 3\,mm transitions, and as the predicted number of detections assuming the corresponding CO LFs from 3\,mm at similar redshifts, and the large velocity gradient modeling of the CO spectral energy distribution from \citet{boogaard20}.
Blue boxes: the probability distribution scales as the observed flux distributions shown in Fig.~\ref{fig_flux_distr}. 
Purple boxes: Constraints on $\rho_{\rm H2}$ based exclusively on the sources detected in the 1.2\,mm dust continuum. 
}
\label{fig_rhoH2_tests}
\end{figure}

Fig.~\ref{fig_rhoH2_tests} compares the $\rho_{\rm H2}$($z$) evolution that results from each assumption (see also Tab.~\ref{tab_rhoH2_tests}). To first order, the $\rho_{\rm H2}$ evolution is unaffected by our treatment of the sources without a counterpart in the catalog. The spread between the $\rho_{\rm H2}$ estimates is most prominent at $z\lsim 0.5$ as a result of low number statistics. Discrepancies are always well within the uncertainties. The main offset comes from restricting our analysis to the secure sources with a 1.2\,mm dust continuum, which typically results in a $\sim 1.5\times$ underestimate of $\rho_{\rm H2}$.

\begin{table}
\caption{\rm The cosmic molecular gas density (mass of molecular gas in galaxies per cosmological volume) as constrained by ASPECS under different methods to identify line candidates without counterparts (see text for details). The quoted values refer to the 1-$\sigma$ upper- and lower condifence boundaries.} \label{tab_rhoH2_tests}
\begin{center}
\begin{tabular}{ccccc}
\hline
 Redshift    & \multicolumn{4}{c}{log $\rho_{\rm H2}$ [\Msun{}\,Mpc$^{-3}$]} \\
             & Forward model & Volume-based & Flux-based & Top sources \\
 (1)    &     (2)       & (3)         & (4)     & (5)      \\
\hline
    0.003---0.369  &   5.171---6.449  &   5.171---6.449  &   5.177---6.453  &		       \\
    0.271---0.631  &   6.196---7.184  &   6.537---7.312  &   6.522---7.240  &	 5.911---7.194 \\
    0.695---1.174  &   7.617---7.957  &   7.278---7.819  &   7.398---7.871  &	 7.138---7.772 \\
    1.006---1.738  &   7.608---7.874  &   7.608---7.874  &   7.769---7.984  &	 7.481---7.816 \\
    2.008---3.107  &   7.266---7.647  &   7.266---7.647  &   7.322---7.663  &	 6.961---7.595 \\
    3.011---4.475  &   7.227---7.517  &   7.227---7.517  &   6.720---7.190  &	 6.766---7.235 \\
\hline
\end{tabular}
\end{center}
\end{table}

\label{lastpage}

\end{document}